\documentclass[aps,prd,superscriptaddress,nofootinbib,11pt]{revtex4}
\usepackage[english]{babel}
\usepackage[utf8]{inputenc}
\usepackage{graphicx}   
\usepackage{slashed}
\usepackage{epstopdf}
\usepackage{verbatim}   
\usepackage{color}      
\usepackage{subfigure}  
\usepackage{multirow}
\usepackage{hyperref}   
\usepackage{float}
\usepackage{epsfig,rotating}
\usepackage{amsmath,amssymb}
\usepackage{dsfont}
\usepackage{slashed}
\restylefloat{table}
\numberwithin{equation}{section}

\newcommand{\vx}{\vec{x}}

\newcommand{\vk}{\vec{k}}

\newcommand{\be}{\begin{equation}}
\newcommand{\ee}{\end{equation}}
\newcommand{\bea}{\begin{eqnarray}}
\newcommand{\eea}{\end{eqnarray}}

\newcommand{\ket}[1]{|#1\rangle}
\newcommand{\bra}[1]{\langle#1|}

\newcommand{\M}{\mathcal M}

\newcommand{\C}{\mathcal{C}}
\newcommand{\R}{\mathcal{R}}
\newcommand{\J}{\mathcal{J}}
\newcommand{\D}{\mathcal{D}}
\newcommand{\walpha}{\widetilde{\alpha}}
\newcommand{\wbeta}{\widetilde{\beta}}
\newcommand{\wgamma}{\widetilde{\gamma}}
\newcommand{\wS}{\widetilde{\Sigma}}
\newcommand{\wR}{\widetilde{\mathbb{R}}}
\newcommand{\wD}{\widetilde{D}}
\newcommand{\wP}{\widetilde{P}}
\newcommand{\oalpha}{\overline{\alpha}}
\newcommand{\obeta}{\overline{\beta}}
\newcommand{\ogamma}{\overline{\gamma}}
\begin{document}
\title{ Effective field theory of particle mixing.}

\author{Shuyang Cao}
\email{shuyang.cao@pitt.edu} \affiliation{Department of Physics, University of Pittsburgh, Pittsburgh, PA 15260}
\author{Daniel Boyanovsky}
\email{boyan@pitt.edu} \affiliation{Department of Physics, University of Pittsburgh, Pittsburgh, PA 15260}

 \date{\today}

\begin{abstract}

We introduce an effective field theory   to study \emph{indirect}   mixing of two fields induced by their couplings to  a common decay channel in a medium. The extension of the method of Lee, Oehme and Yang, the cornerstone of analysis of CP violation in flavored mesons,    to include mixing of   particles  with different masses provides a guide to and benchmark for the effective field theory. The analysis reveals subtle caveats in the description of mixing in terms of the widely used  non-Hermitian effective Hamiltonian, more acute in the non-degenerate case. The effective field theory describes the dynamics of field mixing where the common intermediate states populate a bath in thermal equilibrium, as an \emph{open quantum system}. We obtain the effective action up to second order in the couplings, where indirect mixing is a consequence of  off-diagonal self-energy components. We find that if only one of the mixing fields features an initial expectation value, indirect mixing induces an expectation value of the other field. The equal time two point correlation functions exhibit asymptotic approach to a stationary thermal state, and  the emergence of long-lived \emph{bath induced} coherence which display quantum beats as a consequence of interference of quasinormal modes in the medium.  The amplitudes of the quantum beats are resonantly enhanced in the nearly degenerate case with potential observational consequences.

\end{abstract}


\maketitle

\section{Introduction}\label{sec:intro}
The  dynamics of particle  mixing induced by their coupling to  a common intermediate state or decay channel  is of   broad   fundamental   interest  within the context of CP violation and/or baryogenesis.  Field mixing may also be a consequence of  ``portals'', connecting standard model degrees of freedom to hypothetical ones   via mediator particles beyond the standard model. Such ``portals'' may lead to mixing between fields on different sectors of the ``portal'' via the exchange of these mediators, namely a common intermediate state to which fields on different sides of the portals couple.

Axions,  CP-odd pseudoscalar particles proposed in extensions beyond the standard model as a possible solution of the strong CP problem in Quantum Chromodynamics (QCD)\cite{PQ,weinaxion,wil} could be  a compelling cold dark matter candidate\cite{pres,abbott,dine}. However, various extensions beyond the standard model can include axion-like particles with properties similar to the QCD axion which may also be suitable dark matter candidates\cite{banks,ringwald,marsh,sikivie1,sikivie2}. Just like the QCD axion these axion-like particles couple to photons and gluons via Chern-Simons terms such as $\vec{E}\cdot \vec{B}$ in the case of photons,  or $\widetilde{G}^{\mu\nu;b}G_{\mu \nu;b}$ in the case of gluons,  as a consequence of the chiral anomaly. Their mutual coupling to photons and gluons entails that the various ``flavors'' of axions or axion-like particles may mix via a common intermediate state of photons and gluons. For example processes such as $A \leftrightarrow \gamma \gamma  \leftrightarrow A'$, with $A,A'$ being different axion-like particles, yield  off diagonal self-energy components $\Sigma_{A,A'}$, hence an \emph{indirect} mixing via the common intermediate state.

 A paradigmatic example   in \emph{vacuum} is the mixing of $K^0-\overline{K}^0$ or flavored meson-antimesons as a consequence of a common intermediate states of two or three pions (or the weak interaction box diagram), providing  dynamical observational signatures of CP violation\cite{cpv1,cpv2,cprev,pdg,pila1,pila2}.

Field mixing via a common intermediate state in a thermal medium has recently been studied\cite{mesonmix} within the context of axion-neutral pion mixing after the QCD phase transition, since the   neutral pion couples to two photons precisely via a $U(1)$ Chern Simons term as a consequence of the chiral anomaly.

Recently, it has been realized that topological materials and/or Weyl semimetals also feature emergent axion-like quasiparticles as collective excitations, which couple to electromagnetism via processes akin to the U(1) anomaly\cite{wilczekaxion,wang,narang,nomura,gooth,yu,wilczek,baly,binek,mottola}. Therefore, these ``synthetic'' axions may mix with the cosmological axion in the same manner as pions or generic axion-like particles    in the early Universe. This possibility   motivates the study  of mixing between the cosmological and the ``synthetic'' axions, which  may yield   alternative experimental avenues to probe cosmological axions with condensed matter experiments.

\vspace{1mm}

\textbf{Motivations and  objectives:} Motivated by the ubiquity of     field mixing and its broad relevance in particle physics, cosmology and possibly in condensed matter physics,   we extend the preliminary study of ref.\cite{mesonmix}   and develop a more general  effective field theory framework to study mixing as a consequence of coupling of different fields to common intermediate states or decay channels.  We distinguish \emph{direct} mixing as a result of explicit mixing terms in the Lagrangian, such as off-diagonal mass matrices or kinetic mixing terms, from the \emph{indirect} mixing via common intermediate states leading to off-diagonal self-energy components, such as flavored meson mixing, for example $K^0-\overline{K}^0$. Our study is focused on this latter,  \emph{indirect} mixing case.

The theory of $K^0-\overline{K}^0$ mixing via weak interaction intermediate states was advanced by Lee, Oehme and Yang\cite{lee}, in their pioneering study of CP violation. It is based on the theory of atomic linewidths developed  by Weisskopf and Wigner\cite{ww,sachs,cppuzzle,chiu}, and is the cornerstone of the analysis of mixing dynamics of flavored mesons and CP violation\cite{cpv1,cpv2,cprev,pdg} in terms of an effective non-hermitian Hamiltonian.

Our main focus is to develop an effective field theory framework to study the dynamics of \emph{indirect} mixing  when the particles in the intermediate states are components of a thermal bath as is the case in cosmology. An advantage of the effective field theory formulation of mixing is that it allows to obtain correlation functions in the medium, to understand their approach to thermalization, and the emergence  of long lived coherence, namely off diagonal components of the two point field correlation function that survives in the long time limit even when initially the different fields are uncorrelated.

 The preliminary study of    ref.\cite{mesonmix} focused on  the particular case of axion-neutral pion mixing near the QCD phase transition
  where the axion was assumed to be a light or ultralight CP-odd scalar. In this case there is a  large mass difference between the mixing partners leading to suppression of interference effects. Furthermore,  axions and neutral pions couple to photons with the same operator ($\vec{E}\cdot \vec{B}$) but with different couplings, making this a particular case.

  Instead, here   we   contemplate more general scenarios including that of degenerate or nearly degenerate mixing fields and coupling to intermediate states with  different operators  with non vanishing correlations  in the thermal bath, thereby leading to mixing via off-diagonal self-energy matrix elements. This more general situation may be relevant for CP violation in the early Universe and  yields     far richer dynamics including non-perturbative interference phenomena    in the form of quantum beats that plays an important role in the approach to thermalization and the dynamics of coherence, with possible observational consequences.

  Unlike the case of direct mixing, such as neutrino mixing via an off-diagonal mass matrix, or kinetic mixing, \emph{indirect} mixing in a medium, as is relevant in cosmology, and to the best of our knowledge has not yet been studied at a deeper level.

  Our objectives are \textbf{i:)} to provide a consistent effective field theory framework to study the \emph{dynamics} of mixing via intermediate states in in equilibrium a medium. \textbf{ii:)} to apply this formulation to study the non-equilibrium dynamics of expectation values and correlation functions of the mixing fields, \textbf{iii:)} to focus  in particular  on the approach to thermalization and the emergence  and long time survival of coherence even when initially the mixing fields are uncorrelated.

  The equations of motion obtained from the effective field theory  allow to study the dynamical evolution of expectation values and correlation functions and the emergence and evolution of coherence, hence providing an approach to the study of coherence that complements the quantum master equation\cite{pilamaster,mario,caoqme}.  We also recognize that the effective field theory approach to mixing may also be extended to the case of neutrinos in the mass basis, and may provide an alternative framework to study the quantum kinetics of massive neutrinos in the medium\cite{kainu}. More recently a quantum field theoretical approach to a Boltzmann equation for axions consistently including misaligned condensates has been introduced in ref.\cite{aimarsh}. The formulation of an effective field theory of mixing developed in this study may provide a complementary approach when different types of axions mixing indirectly via  a common intermediate state are considered.

  In this article our main objective is to develop the theoretical framework in general, without specifying particular models or applications, which will be the subject of future study.

  \vspace{1mm}

  \textbf{Brief summary of results:}

  As a prelude to developing the effective field theory framework, in section (\ref{sec:loy}) we extend the Lee-Oehme-Yang (LOY) theory of mixing to the case of non-degenerate mixing particles and with generic couplings to intermediate states,  and solve exactly the equations for the amplitudes, which goes beyond the usual approach based on a non-hermitian effective Hamiltonian\cite{cpv1,cpv2,cprev,pdg}. The generalization to the nearly degenerate and non-degenerate   cases provides an extension to analyze the dynamics of mixing relaxing  the assumption of validity of CPT.  This study serves as a guide and benchmark    towards establishing the effective field theory framework, and also reveals interesting caveats of the usual approach with a non-hermitian Hamiltonian, which become more important in the non-degenerate case and may be  relevant in precision measurements of CP violation. Appendix (\ref{app:single}) discusses the origin of  some of these caveats in the case of a single species.

  In section (\ref{sec:effaction}) we consider \emph{indirect} mixing of two bosonic fields induced by  their couplings to a common decay channel in the medium. These common intermediate states populate a bath in thermal equilibrium.

  We generalize the methods of refs.\cite{shuyang,mesonmix} to obtain the effective action in the in-in or Scwhinger-Keldysh formulation of non-equilibrium quantum field theory\cite{schwinger,keldysh,maha,beilok,feyver} up to second order in couplings.  This effective action determines the time evolution of the reduced density matrix upon tracing the bath degrees of freedom and it describes the dynamics of mixing as an open quantum system. The equations of motion obtained from the effective action are stochastic with self-energy and noise kernels obeying a generalized fluctuation dissipation relation. Indirect mixing is a consequence of off-diagonal self-energy components arising from the correlations of the coupling operators in the bath. The solution of  the equations of motion yield  the time evolution of expectation values and correlation functions in terms of superpositions  of quasinormal modes in the medium. The cases of non-degenerate and nearly degenerate fields are studied in detail. We find that if only one of the fields has an initial non-vanishing expectation value, indirect mixing induces an expectation value for the other field. Furthermore, the equal time two points correlation function approaches a stationary thermal state independent of the initial conditions and even when initially the fields are uncorrelated exhibit an emergent long-lived \emph{bath-induced} coherence, namely off diagonal components. Both diagonal and off-diagonal correlation functions display \emph{quantum beats}, as a consequence of interference of quasinormal modes. The amplitudes of the quantum beats is resonantly enhanced in the case of nearly degenerate fields.   In this section we establish the correspondence between the (LOY) formulation of particle mixing and the effective field theory of mixing.

  Several appendices supplement technical details. Appendix (\ref{app:single}), discusses the caveats associated with a non-hermitian Hamiltonian for a single species. Section (\ref{sec:conclusions}) summarizes the main results and conclusions.

\section{ The Lee-Oehme-Yang (LOY) theory of mixing. \cite{lee,ww}:}\label{sec:loy}

We begin by extending and generalizing the formulation of meson mixing pioneered by Lee, Ohme and Yang\cite{lee}   to analyze  CP violation  in the Kaon system,   which based on the Weisskopf-Wigner theory of atomic linewidths\cite{ww}, to the case when particles of different masses mix via a common set of intermediate states, or common decay channel. Such a generalization will lead us to the formulation of an effective quantum field theory of mixing including the case when the particles in the intermediate states constitute a medium as is relevant in  cosmology.

Consider a system whose Hamiltonian $H$ is given as a soluble part $H_0$ and a perturbation $H_I$: $H=H_0+H_I$. The time evolution of states in the interaction picture
of $H_0$ is given by
\be i \frac{d}{dt}|\Psi(t)\rangle_I  = H_I(t)\,|\Psi(t)\rangle_I,  \label{intpic}\ee
where the interaction Hamiltonian in the interaction picture is
\be H_I(t) = e^{iH_0\,t} H_I e^{-iH_0\,t} \label{HIoft}\,,\ee where $H_I$ is proportional to a set of couplings assumed to be small.

Eqn. (\ref{intpic})  has the formal solution
\be |\Psi(t)\rangle_I = U(t,t_0) |\Psi(t_0)\rangle_I \label{sol}\ee
where   the time evolution operator in the interaction picture $U(t,t_0)$ obeys \be i \frac{d}{dt}U(t,t_0)  = H_I(t)U(t,t_0)\,. \label{Ut}\ee

Now we can expand \be |\Psi(t)\rangle_I = \sum_n C_n(t) |n\rangle \label{decom}\ee where $|n\rangle$ form a complete set of orthonormal states chosen to be eigenfunctions of $H_0$, namely $H_0\ket{n} = E_n\ket{n}$; in the quantum field theory case these are  many-particle Fock states. From eqn.(\ref{intpic}), and the expansion (\ref{decom})  one finds the   equation of motion for the coefficients $C_n(t)$, namely

\be \dot{C}_n(t) = -i \sum_m C_m(t) \langle n|H_I(t)|m\rangle \,. \label{eofm}\ee

Although this equation is exact, it generates an infinite hierarchy of simultaneous equations when the Hilbert space of states spanned by $\{|n\rangle\}$ is infinite dimensional. However, this hierarchy can be truncated by considering the transition between states connected by the interaction Hamiltonian at a given order in $H_I$.

\begin{figure}[ht]
\includegraphics[height=3.0in,width=3.0in,keepaspectratio=true]{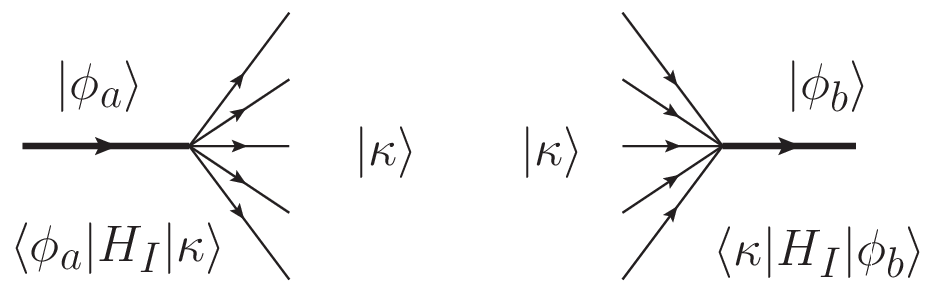}
\caption{Mixing between $\ket{\phi}_a,\ket{\phi}_b$ mediated by a common set of intermediate states $\ket{\kappa}$ . }
\label{fig:transitions}
\end{figure}

Let us consider quantum states $\ket{\phi_1},\ket{\phi_2}$ associated with the meson fields $\phi_{1,2}$ respectively, these may be single particle momentum eigenstates of the Fock quanta of these fields, and focus on  the case when the interaction Hamiltonian does not couple \emph{directly} the states $\ket{\phi_1},\ket{\phi_2}$, namely $\langle\phi_a|H_I|\phi_{b}\rangle =0$. Instead these states are connected to a common set of intermediate states $|\{\kappa\}\rangle$ by  $H_I$, namely $\ket{\phi_{1,2}} \leftrightarrow |\{\kappa\}\rangle \neq \ket{\phi_{1,2}}$, as depicted in fig. (\ref{fig:transitions}).

 The states $\ket{\phi_1},\ket{\phi_2}$ mix as a consequence of this indirect coupling through the common set of intermediate states, namely $\ket{\phi_{1,2}} \leftrightarrow |\{\kappa\}\rangle \leftrightarrow \ket{\phi_{1,2}}$, yielding an off-diagonal self-energy matrix.  If $H_I$ has non-vanishing matrix elements $\langle \phi_i|H_I|\phi_j\rangle \neq 0$, we assume that these have been absorbed into terms in $H_0$ and only consider  transitions between $\ket{\phi_i}$ and other states $\ket{\kappa} \neq \ket{\phi_{1,2}}$ mediated by $H_I$.

In the   subspace $\ket{\phi_1},\ket{\phi_2},  |\{\kappa\}\rangle$ the quantum state in the interaction picture is given by
\be |\Psi\rangle_I(t) = C_1(t)\ket{\phi_1}+ C_2(t)\ket{\phi_2}+ \sum_{\{\kappa\}} C_{\kappa}(t) | \kappa \rangle \,, \label{state}\ee
and the set of equations (\ref{eofm})   become
\bea
\dot{C}_{1}(t) & = & -i \sum_{\{\kappa\}}  \langle\phi_1|H_I(t)| \kappa \rangle \, C_{\kappa}(t)\, \label{eqnc1} \\
\dot{C}_{2}(t) & = & -i \sum_{\{\kappa\}}  \langle\phi_2|H_I(t)| \kappa \rangle \, C_{\kappa}(t)\,, \label{eqnc2} \\
\dot{C}_{ \kappa}(t) & = & -i  \Bigg[\langle  \kappa |H_I(t)|\phi_1\rangle\, C_{ 1}(t)+ \langle  \kappa |H_I(t)|\phi_2\rangle\,  C_{ 2}(t)\Bigg]\,.\label{eqncm}\eea   where   the time dependent transition matrix elements are given by
\be \langle l|H_I(t)|m\rangle = T_{lm}\,  e^{i(E_l-E_m)t} ~~;~~ T_{lm}= \langle l|H_I(0)|m\rangle \,,\label{mtxele} \ee hermiticity of $H_I$ entails that
  \be T_{ml} = T^*_{lm} \,.\label{hermi}\ee

  The set of equations (\ref{eqnc1}-\ref{eqncm}), truncates the hierarchy of equations by neglecting the transitions between the states $|\{\kappa\}\rangle$ and  $|\{\kappa'\}\rangle \neq |\{\kappa\}\rangle,\ket{\phi_{1,2}}$, such transitions connect the states $\ket{\phi_{1,2}} \leftrightarrow |\{\kappa'\}\rangle $ at higher order in $H_I$ and are neglected up to $\mathcal{O}(H^2_I)$.  Truncating the hierarchy closes the set of equations for the amplitudes, effectively reducing the set of states to a closed subset in the full Hilbert space. As a familiar example, let us consider the   case where  $\ket{\phi_{1,2}}$ correspond to $K^0,\overline{K}^0$ mesons mixing via a common decay channel into two pions (there is also the three pion decay channel), so that $K^0 \leftrightarrow 2\pi \leftrightarrow \overline{K}^0$.

  Taking the normalized initial  quantum state $ \ket{\Psi(t=0)}$ as a coherent linear superposition of the single particle states $\ket{\phi_{1,2}}$, it is given by
\be \ket{\Psi(t=0)} = \Big(C_{1}(0)\ket{\phi_1} + C_{2}(0)\ket{\phi_2}\Big)\otimes \ket{0_\kappa}  \,,\label{inistate} \ee where $\ket{0_\kappa}$ is the vacuum state for the intermediate states $\ket{\kappa}$, corresponding to setting
\be C_{\kappa}(0) = 0 \,, \label{cmini}\ee for the excited $\ket{\kappa}$ states, and with normalization condition
\be   |C_{1}(0)|^2+ |C_{2}(0)|^2  =1 \,. \label{unit}  \ee

\vspace{1mm}

\textbf{Unitarity:} The set of equations (\ref{eqnc1}-\ref{eqncm}) describe \emph{unitary time evolution} in the restricted Hilbert space of states $\ket{\phi_1},\ket{\phi_2},|\kappa\rangle$ which is a sub-set of the full Hilbert space of the theory that is closed under the equations  of motion (\ref{eqnc1}-\ref{eqncm}). Unitarity can be seen as follows: using the equations (\ref{eqnc1}-\ref{eqncm}), and noticing that $\langle l|H_I(t)|m\rangle^* = \langle m|H_I(t)|l\rangle $ because $H_I(t)$ is an Hermitian operator, it follows that
\be \frac{d}{dt} \Bigg[ | {C}_{1}(t)|^2+   | {C}_{2}(t)|^2 + \sum_{\{\kappa\}} | {C}_{\kappa}(t)|^2\Bigg] =0 \,,\label{totder}\ee and the initial conditions (\ref{unit},\ref{cmini}) yield \be |{C}_{1}(t)|^2+   | {C}_{2}(t)|^2 + \sum_{\{\kappa\}} | {C}_{\kappa}(t)|^2 = 1 \,.  \label{unitarity} \ee This is the statement that time evolution within the sub-Hilbert space $\Big\{\ket{\phi_1},\ket{\phi_2},|\kappa\rangle\Big\}$  is unitary.

In particular if the $\phi_{1,2}$ states decay, it follows that $|C_{1,2}(t=\infty)|^2=0$,  and
\be \sum_{\kappa}|C_{\kappa}(t=\infty)|^2 = 1 \,. \label{inftyti}\ee

The set of equations (\ref{eqncm}) with the initial condition (\ref{cmini}) can be integrated to yield
\be C_\kappa(t) = -i \int^t_0 \Bigg[T_{\kappa 1} \,e^{i(E_\kappa-E_1)t'}\, C_{ 1}(t')+ T_{\kappa 2} \,e^{i(E_\kappa-E_2)t'}\,  C_{ 2}(t')\Bigg]\,dt' \,,\label{cmoft} \ee where the labels $1,2$ correspond to $\phi_{1,2}$. Upon inserting the solution  (\ref{cmoft}) into the equations (\ref{eqnc1},\ref{eqnc2}) lead to the following set of equations for the coefficients $C_{1}(t),C_{2}(t)$
\be
  \dot{C}_{1}(t)   =  -  \int^t_0  \sum_{\kappa}\Bigg\{   |T_{1\kappa}|^2 \, e^{i(E_1-E_\kappa)(t-t')}\,   {C}_{1}(t') +  T_{1\kappa}T_{\kappa 2}\,e^{i(E_1-E_2)t}\,e^{i(E_2-E_\kappa)(t-t')} {C}_{2}(t') \Bigg\} \, dt'\,, \label{dotC1} \ee
\be \dot{C}_{2}(t)   =   -  \int^t_0  \sum_{\kappa}\Bigg\{  T_{2\kappa}T_{\kappa 1}\,e^{i(E_2-E_1) t}\,e^{i(E_1-E_\kappa)(t-t')} {C}_{1}(t') + |T_{2\kappa}|^2\,e^{i(E_2-E_\kappa)(t-t')}\,    {C}_{2}(t') \Bigg\}  \, dt' \,. \label{dotC2}
\ee

This procedure of solving for the amplitudes of  the intermediate states   plays the role of ``integrating out'' or ``tracing over'' the $\kappa$ degrees of freedom,   yielding an effective set of equations of motion for the amplitudes of the single particle states $\ket{\phi_{1,2}}$. Since the interaction Hamiltonian $H_I$ is assumed to include a weak coupling, the amplitude equations (\ref{dotC1},\ref{dotC2}) are exact up to second order in this coupling. Pictorially, this procedure is equivalent to joining the legs representing the $\chi$ field together in fig. (\ref{fig:transitions}), thereby forming a loop or loops that yield(s) the self-energy.  

\subsection{Exact solutions:}\label{subsec:exact}

The set of amplitude equations (\ref{dotC1},\ref{dotC2}), can be solved exactly. For this purpose it is convenient to define
\be e^{-iE_1 t}\,C_1(t)\equiv A_1(t) ~~;~~ e^{-iE_2 t}\,C_2(t)\equiv A_2(t)\,,\label{As}\ee and to  introduce  the spectral densities
\be \mu_{ab}(k_0) =  \,\sum_{\kappa} T_{a \kappa} T_{\kappa b} \,\delta(k_0-E_{\kappa}) = \mu^*_{ba}(k_0)~~;~~ a,b = 1,2 \,, \label{rhosd}\ee where the second identity follows from equation (\ref{hermi}). The set of equations for the amplitudes $A_{1,2}$, following from (\ref{dotC1},\ref{dotC2}) are written more compactly by introducing the self-energy matrix
\be \sigma_{ab}(t-t') = \int^{\infty}_{-\infty} \mu_{ab}(k_0)\,e^{-ik_0(t-t')}\, {dk_0}
\,. \label{sigmasd}\ee This self-energy has an intuitive interpretation as a second order Feynman diagram wherein the lines representing the intermediate states $\ket{\kappa}$ in fig. (\ref{fig:transitions}) are joined into ``propagators'' yielding a (multi)loop diagram,   representing the self-energy up to second order in $H_I$.

In terms of the self-energy the set of equations (\ref{dotC1},\ref{dotC2}) become
\be \dot{A}_a(t) +i E_a \, A_a(t)+ \int^t_0 dt'   \,\sum_{  b}\sigma_{ab}(t-t')\, A_b(t')=0\,. \label{equafin}\ee   This equation can be solved via a Laplace transform. Defining the Laplace transforms for $\mathrm{Re}(s)>0$
\be \widetilde{A}_a(s) = \int^\infty_0 e^{-st}\,A_a(t)\,dt~~;~~ \widetilde{\sigma}_{ab}(s) = \int^\infty_0 e^{-st}\,\sigma_{ab}(t)\,dt = \int^{\infty}_{-\infty} \, \frac{\mu_{ab}(k_0)}{s+ik_0} \, {dk_0} \,,\label{laplas} \ee the set of equations (\ref{equafin}) leads to
\be  \Bigg(
                   \begin{array}{cc}
                     M_{11} & M_{12}\\
                     M_{21} & M_{22} \\
                   \end{array}
                 \Bigg)\, \Bigg(\begin{array}{c}
                   \widetilde{A}_1(s) \\
                   \widetilde{A}_2(s)
                 \end{array}\Bigg) =  \Bigg(\begin{array}{c}
                    {A}_1(0) \\
                    {A}_2(0)
                 \end{array}\Bigg)\,, \label{matxlapla} \ee
 with matrix elements
\bea  {M}_{11} & = & s+i\,E_1+\widetilde{\sigma}_{11}(s)\,, \label{m11} \\
  {M}_{12} & = &  \widetilde{\sigma}_{12}(s)~~;~~  {M}_{21}  =   \widetilde{\sigma}_{21}(s)\,, \label{moffd} \\
  {M}_{22} & = & s+i\,E_2+\widetilde{\sigma}_{22}(s)\,, \label{m22}  \eea   where we suppressed the dependence of the matrix elements $M_{ij}$ on the Laplace variable $s$ to simplify notation but it is implicit in all matrix elements.

It proves convenient to introduce
\bea \overline{M} & = &  \frac{1}{2}\,\big(M_{11}+M_{22} \big)\,,\label{mbar}\\D & = &  \Big [\big(M_{11}-M_{22} \big)^2+ 4\,M_{12}M_{21} \Big]^{1/2}\,,\label{D}\\
\alpha & = & \frac{M_{11}-M_{22}}{D}~~;~~\beta = \frac{2\,M_{12}}{D}~~;~~\gamma = \frac{2\,M_{21}}{D}\,,\label{abg}
\eea where $\alpha,\beta,\gamma$ fulfill the relation
\be \alpha^2+\beta\gamma =1 \,.\label{abgrela} \ee  The inverse of the matrix with elements $M_{ab}$   yields the Laplace Green's function , which is given by (see appendix (\ref{app:green}) for details)
\be \Bigg(\begin{array}{c}
                   \widetilde{A}_1(s) \\
                   \widetilde{A}_2(s)
                 \end{array}\Bigg) = \Bigg[  \frac{\mathbb{P}_-(s)}{\overline{M} (s) - \frac{ D(s)}{2} } +  \frac{\mathbb{P}_+(s)}{\overline{M} (s) + \frac{ D(s)}{2} }  \Bigg]\, \Bigg(\begin{array}{c}
                    {A}_1(0) \\
                    {A}_2(0)
                 \end{array}\Bigg)\,,  \ee  where the projector operators (see appendix (\ref{app:green})) are given by
                 \be \mathbb{P}_\pm (s)  =    \frac{1}{2} \Big(\mathbf{1}\pm \mathbb{R}(s) \Big) ~~;~~ \mathbb{R}(s) = \left(
                                                                           \begin{array}{cc}
                                                                             \alpha(s) & \beta(s) \\
                                                                             \gamma(s) & -\alpha(s) \\
                                                                           \end{array}
                                                                         \right)  \,. \label{projs}\ee Finally the time evolution of the amplitudes $A_{1,2}(t)$ is
obtained via the inverse Laplace transform,
\be \Bigg(\begin{array}{c}
     A_{1}(t) \\
      A_{2}(t)
    \end{array}\Bigg) = \int_{\mathcal{C}}\, e^{st}\,  \Bigg[  \frac{\mathbb{P}_-(s)}{\overline{M} (s) - \frac{ D(s)}{2} } +  \frac{\mathbb{P}_+(s)}{\overline{M} (s) + \frac{ D(s)}{2} }  \Bigg]\, \,\frac{ds}{2\pi i}\, \Bigg(\begin{array}{c}
                    {A}_1(0) \\
                    {A}_2(0)
                 \end{array}\Bigg) \,,\label{iLT}\ee where the Bromwich countour $\mathcal{C}$ runs parallel to the imaginary axis to the right of all the singularities in the complex $s$ plane. Stability implies that the real part of the singularities are negative, therefore the contour corresponds to $s = i\nu+ \epsilon~,~  -\infty \leq \nu \leq \infty ~,~ \epsilon \rightarrow 0^+$. It is convenient to change variables to $\nu = -\omega$, in terms of  which
\be  \widetilde{\sigma}_{ab}(s=i(-\omega-i\epsilon))\equiv i\Delta_{ab}(\omega) = i \,\Bigg[ \int^\infty_{-\infty}\,\mathcal{P}\,\Bigg( \frac{\mu_{ab}(k_0)}{\omega - k_0}\Bigg)\,dk_0 - i\,\pi\,\mu_{ab}(\omega)\Bigg]\,, \label{sigofnu}\ee where $\mathcal{P}$ stands for the principal part. The relation (\ref{rhosd}) implies that
\be \Delta_{ba}(\omega) =   \,\Bigg[ \int^\infty_{-\infty}\,\mathcal{P}\,\Bigg( \frac{\mu^*_{ab}(k_0)}{\omega - k_0}\Bigg)\,dk_0 - i\,\pi\,\mu^*_{ab}(\omega)\Bigg]\,. \label{sigofnucc}\ee

Upon this analytic continuation, equation (\ref{iLT}) becomes
\be \Bigg(\begin{array}{c}
      A_{1}(t) \\
      A_{2}(t)
    \end{array}\Bigg) = -\int^\infty_{-\infty} \, e^{-i\omega t}\,  \Bigg[  \frac{\mathbb{P}_-(\omega)}{\omega-W_-(\omega) } +  \frac{\mathbb{P}_+(\omega)}{ \omega- W_+(\omega) }  \Bigg]\, \,\frac{d\omega}{2\pi i}\, \Bigg(\begin{array}{c}
                    {A}_1(0) \\
                    {A}_2(0)
                 \end{array}\Bigg) \,,\label{iLTw}\ee where
\bea W_{\pm}(\omega) & = &  \frac{1}{2}\Bigg\{\Big( E_1+E_2+\Delta_{11}(\omega)+\Delta_{22}(\omega) \Big)\pm \, D(\omega)\,  \Bigg\}\,,\label{Wpm} \\ D(\omega) &= &  \Big[(E_1-E_2+\Delta_{11}(\omega)-\Delta_{22}(\omega) )^2+4\Delta_{12}(\omega)\Delta_{21}(\omega) \Big]^{1/2}\,,\label{dofw} \eea and $\mathbb{P}_{\pm}(\omega)$ are the analytic continuation of $\mathbb{P}_{\pm}(s)$ for $s\rightarrow -i\omega+\epsilon$.

The bracket inside the integral in (\ref{iLTw}) has a simple interpretation, it is the Dyson (geometric) resummation of the second order self-energy   matrix, the time evolution obtained from (\ref{iLTw}) includes this resummation of second order self-energy corrections. Note that as a consequence of the projector matrices being off-diagonal, even when one of the amplitudes vanishes initially, for example if $A_2(0)=C_2(0)=0$, it becomes  non-vanishing at a later time. This observation will have interesting implications in the analysis of in-medium mixing in the next section.

In the weak coupling limit we invoke the Breit-Wigner approximation, valid in the intermediate time regime,  where each term in (\ref{iLTw}) features a complex pole in the lower half $\omega$ plane at\footnote{More precisely, the poles are in the second Riemann sheet, but close to the real axis in the complex $\omega$-plane.}
\be \omega_{\pm} = W_{\pm}(\omega_\pm) \equiv \varepsilon_{\pm} - i \frac{\Gamma_{\pm}}{2} \,,\label{bwpoles}\ee where $\varepsilon_{\pm}$ are the renormalized frequencies and $\Gamma_{\pm}$ the decay rates. In weak coupling, it follows that $\Gamma_{\pm} \propto H^2_I\ll E_{1,2}$, we will refer to these complex frequencies as \emph{quasinormal modes}. For a vanishing damping rate, these are the usual normal modes associated with the coupling of harmonic oscillators, the ``\emph{quasi}'' reflects their damping as a consequence of their coupling to and decay into a (common) continuum.

Evaluating (\ref{iLTw}) by contour integration closing in the lower half $\omega$ plane for $t>0$, and expanding near the complex poles $W_{\pm}(\omega) = W_{\pm}(\omega_\pm)+(\omega - \omega_{\pm})\,dW_\pm(\omega)/d\omega|_{\omega = \omega_{\pm}}+ \cdots$,  we obtain the final result
\be   \Bigg(\begin{array}{c}
      A_{1}(t) \\
      A_{2}(t)
    \end{array}\Bigg) = \Bigg[e^{-i\omega_+ t} \,Z_+\, \mathbb{P}_+(\omega_+)+ e^{-i\omega_- t} \,Z_-\, \mathbb{P}_-(\omega_-)\Bigg]\, \Bigg(\begin{array}{c}
                    {A}_1(0) \\
                    {A}_2(0)
                 \end{array}\Bigg)\,, \label{solLT}\ee where $Z_\pm = \big[1- dW_{\pm}(\omega)/d\omega|_{\omega_\pm}\big]^{-1}$, and
 \be \mathbb{P}_{\pm}(\omega_\pm) = \frac{1}{2}  \left(
                                                                           \begin{array}{cc}
                                                                            1\pm \alpha(\omega_\pm) & \pm \,\beta(\omega_\pm) \\
                                                                             \pm\,\gamma(\omega_\pm)& 1\mp \alpha(\omega_\pm) \\
                                                                           \end{array}
                                                                         \right)\,,\label{ppmlast} \ee

               \noindent  this result is general.

    The Breit-Wigner approximation relies on weak coupling so that the width of the state is much smaller than its mass and that the distance between the real part of the pole and the beginning of the multiparticle cuts must be much larger than the half-width of the particle. This entails that the spectral representation of the propagator  can be well approximated by a Lorentzian centered at the real part of the pole with the width determined by the imaginary part of the self-energy at the position of the pole. Furthermore, this entails that the spectral density of the self-energy is finite and smooth near the value of the pole. This is the same criterion as in Fermi's golden rule.

               It is important to highlight that the Breit-Wigner approximation leading to the result (\ref{solLT}) is valid only during an intermediate time regime, it is neither valid as $t\rightarrow 0$ nor at very long time, when power law corrections emerge\cite{fonda,maiani,misra,chiu,naka}.

As analyzed in detail in these references, the asymptotic late time behavior of the integral in eqn. (\ref{iLTw}) is determined by the behavior of the spectral density at the threshold of multiparticle cuts which yields a power law   that emerges when the amplitude is already perturbatively small (see ref.\cite{boyatta} for a specific example), and the behavior at  early times, $t\rightarrow 0$ receives contributions from the the full spectral density, contributing to  a renormalization of the amplitude of the field. We refer to the intermediate time scale, as the scales between these two limits that depend specifically on the details of the spectral density of the self-energy. However, as is expected in the case of a weakly coupled theory,  the intermediate time scale in which there is exponential decay is generically wide, and is captured reliably by the usual Breit-Wigner approximation of the propagator.

  Therefore the extrapolation to $t \rightarrow 0$ is not consistent with this approximation.
               In fact, the wave function renormalization is a consequence of ``dressing'' and renormalization during an initial transient time scale  describing the formation of a \emph{quasiparticle}\cite{boyzeno}, in renormalizable theories it is usually ultraviolet divergent. The time scale of formation of the  \emph{quasiparticle} is typically associated with the ultraviolet behavior of the spectral density, it is in general much shorter than the typical oscillation and decay time scales of the particle\cite{boyzeno}.


                  In the following analysis  we assume without loss of generality that $E_1 \geq E_2$,  and  consistently with perturbation theory,   that $E_{1,2} \gg \Delta_{ab}\propto H^2_I$. Furthermore, from the identity (\ref{abgrela}) we choose
\be \alpha(\omega) = \sqrt{1-\beta(\omega)\gamma(\omega)}\,, \label{alfi}\ee
hence, in the limit of vanishing coupling $\Delta_{ab} \rightarrow 0$, it   follows that
\be \omega_+ \rightarrow E_1\,,\,\omega_- \rightarrow E_2\,,\, \alpha \rightarrow 1\,,\,\beta,\gamma \rightarrow 0 \,,\,\mathbb{P}_+ \rightarrow \left(
                                                                                                                       \begin{array}{cc}
                                                                                                                         1 & 0 \\
                                                                                                                         0 & 0 \\
                                                                                                                       \end{array}
                                                                                                                     \right)\,,\, \mathbb{P}_- \rightarrow \left(
                                                                                                                       \begin{array}{cc}
                                                                                                                         0 & 0 \\
                                                                                                                         0 & 1 \\
                                                                                                                       \end{array}
                                                                                                                     \right)\,,
 \label{lims} \ee therefore in this limit   the amplitudes $C_{1,2}$ do not depend on time as it must be the case in absence of interactions.

 Two limits are important: \textbf{i:)} $E_1-E_2 \gg \Delta_{ab}$, to which we refer as the \emph{non-degenerate case} and \textbf{ii:)} $E_1-E_2 \lesssim \Delta_{ab}$, to which we refer as the (nearly) \emph{degenerate} case. The first case describes, for example, the mixing between axion-like particles and a neutral pseudoscalar meson as studied in ref.\cite{mesonmix}, such as the pion, with the pion mass much larger than that of the axion.  The second case includes neutral (pseudoscalar) flavored meson-antimeson mixing, such as $K^0-\overline{K}^0$ under the condition of charge conjugation, parity and time reversal (CPT) invariance (in which case $E_1=E_2~;~\Delta_{11}(s)=\Delta_{22}(s)$). This second case also applies to neutral meson mixing if there is a small (CPT) violation, in which case $E_1, E_2$ and  the diagonal matrix elements $\Delta_{11},\Delta_{22}$ may be slightly different but small compared to the individual energies $E_{1,2}$.

 \vspace{1mm}

  {\textbf{I:) Non-degenerate case: $E_1 - E_2 \gg \Delta_{ab}$}.} In this case we can approximate
 \be  D(\omega) \simeq E_1-E_2+\Delta_{11}(\omega)-\Delta_{22}(\omega)+\mathcal{O}(\Delta^2)\,,\label{nodeg}\ee from which it follows that to leading order ($\mathcal{O}(\Delta)$)
 \be W_+(\omega) = E_1 + \Delta_{11}(\omega)~;~ W_-(\omega) = E_2 + \Delta_{22}(\omega)\,, \label{Wpmnodeg}\ee and to leading order in couplings, the complex poles are at
 \bea \omega_+  & = &  E_1 + \Delta_{11}(E_1) =  E^R_1- i\frac{\Gamma_+}{2}~~;~~ E^R_1 = E_1 + \mathrm{Re}\Delta_{11}(E_1)\,,\, \Gamma_+ =  2\pi\,\rho_{11}(E_1)\,,\label{wpnodeg}\\
 \omega_-  & = & E_2 + \Delta_{22}(E_2)=  E^R_2- i\frac{\Gamma_-}{2}~~;~~ E^R_2 = E_2 + \mathrm{Re}\Delta_{22}(E_2)\,,\, \Gamma_- =  2\pi\,\rho_{22}(E_2)\,,\label{wmnodeg}\eea where $E^R_{1,2}$ are the renormalized energies.  Up to leading order $\mathcal{O}(\Delta)$, it is straightforward to find that  the time dependent amplitudes are given by
 \bea A_1(t) & = &  \,  Z_+ \,A_1(0)\,e^{-i\omega_+ t} +\frac{1}{2}\,\Bigg[  \frac{\Delta_{12}(E_1)\,\,e^{-i\omega_+ t}-\Delta_{12}(E_2)\,\,e^{-i\omega_- t}}{E^R_1-E^R_2} \Bigg]\,A_2(0)  \,,\label{A1tex}\\ A_2(t) & = &    Z_- \,A_2(0)\,e^{-i\omega_- t} +\frac{1}{2}\,\Bigg[  \frac{\Delta_{21}(E_1)\,\,e^{-i\omega_+ t}-\Delta_{21}(E_2)\,\,e^{-i\omega_- t}}{E^R_1-E^R_2} \Bigg]\,\,A_1(0)   \,.\label{A2tex}
 \eea

 The terms in brackets in (\ref{A1tex},\ref{A2tex}) are perturbatively small in this case because $\Delta_{ab} \ll E_1-E_2$. Since $Z_{\pm} \simeq 1 + \mathcal{O}(H^2_I)$ we neglected them in the terms in the brackets, which are already of $\mathcal{O}(\Delta)\propto H^2_I$.

\vspace{1mm}

 {\textbf{II:) (Nearly) degenerate case: $E_{1,2} \gg \Delta_{ab};E_1-E_2 \lesssim \Delta_{ab}$.}} In this case we write
\be \frac{E_1+E_2}{2} \equiv \overline{E}\gg \Delta_{ab} ~;~ E_1-E_2 \equiv \delta \lesssim \mathcal{O}(\Delta)\,,\label{means}\ee and to leading order in $\Delta$ the complex poles are given by
\be \omega_{\pm} =  \overline{E}+ \frac{1}{2}\Big(   \Delta_{11}(\overline{E})+ \Delta_{22}(\overline{E})\Big)\pm \frac{D(\overline{E})}{2} \equiv \varepsilon_{\pm}-i\,\frac{\Gamma_{\pm}}{2} = \overline{E}^R- i \frac{\overline{\Gamma}}{2}\pm \frac{D(\overline{E})}{2}  \,, \label{polesdeg}\ee where
\be \overline{E}^R = \overline{E}+ \frac{1}{2}\Big(  \mathrm{Re}\Delta_{11}(\overline{E})+\mathrm{Re}\Delta_{22}(\overline{E})\Big)~;~ \overline{\Gamma} = - \frac{1}{2}\Big(  \mathrm{Im}\Delta_{11}(\overline{E})+\mathrm{Im}\Delta_{22}(\overline{E})\Big) \,.\label{overEren}\ee From eqns. (\ref{rhosd},\ref{sigofnu}) it follows that $\overline{\Gamma} > 0$. Since in this case $\delta \lesssim \Delta_{ab}$, we find  that $\beta(\omega_\pm) \simeq \gamma(\omega_\pm) \simeq \mathcal{O}(1)$, therefore in this case all matrix elements of the projectors $\mathbb{P}_{\pm} \simeq \mathcal{O}(1)$. However, to leading order in $\Delta$ we find $\beta(\omega_\pm) = \beta(\overline{E})~;~\gamma(\omega_\pm) = \gamma(\overline{E})~;~\alpha(\omega_\pm) = \alpha(\overline{E})$, consequently $\mathbb{P}_\pm(\omega_\pm) =\mathbb{P}_\pm(\overline{E})$.

In this (nearly) degenerate case, the individual energies are much larger than the respective widths and the energy difference is smaller than or of the same order of the
imaginary part of the self-energies evaluated at $(E_1+E_2)/2$. Therefore, in this (nearly) degenerate case, the Breit-Wigner approximation is valid and   we find  to leading order  in $\Delta$ the time dependent amplitudes
\be   \Bigg(\begin{array}{c}
      A_{1}(t) \\
      A_{2}(t)
    \end{array}\Bigg) =  \Bigg[
    \frac{1}{2}\Big(Z_+\,e^{-i\omega_+ t}+Z_-\,e^{-i\omega_- t} \Big)\,\mathbb{I} + \mathbb{R}(\overline{E}) \Big(Z_+\,e^{-i\omega_+ t}-Z_-\,e^{-i\omega_- t} \Big)  \Bigg]\, \Bigg(\begin{array}{c}
                    {A}_1(0) \\
                    {A}_2(0)
                 \end{array}\Bigg)\,, \label{solLTnd}\ee with
                 \be \mathbb{R}(\overline{E})= \frac{1}{2}  \left(
                                                                           \begin{array}{cc}
                                                                             \alpha(\overline{E}) &  \,\beta(\overline{E}) \\
                                                                              \,\gamma(\overline{E})&  -\alpha(\overline{E}) \\
                                                                           \end{array}
                                                                         \right)\,.\label{capRE} \ee  We can now compare this result with the usual result for flavored meson-anti meson mixing, such as $K^0-\overline{K}^0$ under the conditions of
 (CPT) invariance, which implies $E_1=E_2=\overline{E}~;~\Delta_{11}({\overline{E}})=\Delta_{22}(\overline{E})$.       In this case, and for the purpose of comparison, we define
 \be \Delta_{ab}(E) = m_{ab}- i\frac{\Gamma_{ab}}{2} ~~;~~ m_{ab} \equiv \int^\infty_{-\infty}\,\mathcal{P}\,\Bigg( \frac{\mu_{ab}(k_0)}{\overline{E} - k_0}\Bigg)\,dk_0 ~;~ \Gamma_{ab} \equiv  2\pi\,\mu_{ab}(\overline{E}) \,,\label{delabdef2}\ee in terms of which we find
 \bea D(\overline{E}) & = &  2\Bigg[ \Big(m_{12}- i\frac{\Gamma_{12}}{2}\Big)\,\Big(m^*_{12}- i\frac{\Gamma^*_{12}}{2}\Big)\Bigg]^{1/2} \,,\label{Dege}\\
\alpha(\overline{E}) & = & 0 \,,\label{alfadege}\\
\beta(\overline{E}) & = & \Bigg[\frac{m_{12}- i\frac{\Gamma_{12}}{2}}{m^*_{12}- i\frac{\Gamma^*_{12}}{2}} \Bigg]^{1/2}\,,\label{betadege}\\
\gamma(\overline{E}) & = & \Bigg[\frac{m^*_{12}- i\frac{\Gamma^*_{12}}{2}}{m_{12}- i\frac{\Gamma_{12}}{2}} \Bigg]^{1/2}= \frac{1}{\beta(\overline{E})}\,,\label{gamadege} \eea yielding
\bea A_1(t) & = &    \Bigg[\,f_+(t)\,A_1(0)+\beta(\overline{E})\,f_-(t)\,A_2(0)\Bigg]\,\label{A1dege}\\
A_2(t) & = &  \Bigg[\,f_+(t)\,A_2(0)+\gamma(\overline{E})\,f_-(t)\,A_1(0)\Bigg]\label{A2dege}\,,\eea with
\be f_{\pm}(t) = \frac{1}{2}\,\Big(Z_+\,e^{-i\omega_+ t}\pm Z_-\,e^{-i\omega_- t} \Big) \,.\label{fpm}\ee Setting $Z_\pm =1$, the expressions (\ref{A1dege},\ref{A2dege}) with (\ref{fpm}) are the usual ones for the case of flavored meson-antimeson mixing with (CPT) symmetry\cite{cpv1,cpv2,pdg,cppuzzle,cprev}. In reference \cite{cppuzzle} the contribution from wave function renormalization was neglected\footnote{See appendix A, footnote in page 102 in   reference\cite{cppuzzle}, where it is explicitly stated that such contribution was neglected but would modify the amplitudes.} but it was recognized that it would modify the amplitudes. Therefore, with $Z_\pm \simeq 1 + \mathcal{O}(\Delta)$ it is clear  that neglecting the wave function renormalizations affects the amplitudes at second order in the interaction. This perturbative correction \emph{may} become relevant for precision measurements of flavor mixing.

\vspace{1mm}

\subsection{Markov approximation: the effective non-Hermitian Hamiltonian}\label{subsec:markov}

Let us write
\be \int^{t'}_0 \sum_{\kappa} T_{a\kappa}T_{\kappa b}\,  e^{i(E_b-E_{\kappa})(t-t'')}\,dt''   \equiv W_{ab}[t;t']~~;~~ W_{ab}[t,0]=0 \,\label{defW}\ee so that

\be \sum_{\kappa} T_{a\kappa}T_{\kappa b} e^{i(E_b-E_{\kappa})(t-t')}= \frac{d}{dt'}W_{ab}[t,t']\,,\label{iden} \ee inserting this definition in (\ref{dotC1},\ref{dotC2}) and integrating by parts
\be \int^t_0 \frac{d}{dt'}W_{ab}[t,t']\,C_b(t')\,dt' = W_{ab}[t;t]\,C_b(t)-\int^t_0  W_{ab}[t,t']\,\frac{d}{dt'}C_b(t')\,dt'\,,\label{intparts}\ee since $T_{a\kappa} T_{\kappa b} \propto H^2_I$ and from the evolution equations (\ref{dotC1},\ref{dotC2}) it follows that $\dot{C}_a \propto H^2_I$ therefore,  the second term in (\ref{intparts}) is of $\mathcal{O}(H^4_I)$ and will be neglected to leading order in the interaction, namely $H^2_I$.

 Hence, up to $\mathcal{O}(H^2_I)$, the evolution equations for the amplitudes (\ref{dotC1},\ref{dotC2}) become
\bea \dot{C}_1(t) & = & - \Big\{W_{11}[t;t]\,C_1(t)+ e^{i(E_1-E_2)\,t}\,W_{12}[t;t]\,C_2(t)    \Big\}\,\label{dotc1fin} \\
\dot{C}_2(t) & = & - \Big\{e^{i(E_2-E_1)\,t}\,W_{21}[t;t]\,C_1(t)+  W_{22}[t;t]\,C_2(t)    \Big\}\,.\label{dotc2fin}
\eea With the definitions (\ref{As})
  the amplitude equations become
\bea i\,\dot{{A}}_1(t) & = &  E_1 A_1 - i\,W_{11}[t;t]A_1-i\,W_{12}[t;t]A_2 \,\label{amp1} \\
 i\,\dot{{A}}_2(t) & = &  E_2 A_2 - i\,W_{21}[t;t] A_2-i\,W_{22}[t;t]A_2 \,.\label{amp2}\eea With
 \be W_{ab}[t;t] = \sum_{\kappa} T_{a\kappa}T_{\kappa b}\,\int^{t}_0    e^{i(E_b-E_{\kappa})(t-t' )}\,dt'= \int^{t}_0 \int^{\infty}_{-\infty} \mu_{ab}(k_0)\,e^{i(E_b-k_0)(t-t')}\,dk_0 \,dt'\,, \label{Wabtt} \ee where we used the definition of the spectral density, eqn. (\ref{rhosd}). We highlight, that this first step in the Markov approximation is equivalent to the full set of equations consistently up to   order  $H^2_I$,  since the neglected terms of $\mathcal{O}(\Delta^2)\simeq H^4_I$.

 Because in the nearly degenerate case $E_1-E_2 \lesssim \Delta_{ab} \propto H^2_I$, the first stage of the Markov approximation, yielding eqns. (\ref{amp1},\ref{amp2}) is consistent with this case.

 The set of equations (\ref{amp1},\ref{amp2}) can be written in terms of a \emph{time dependent Hamltonian}
 \be i \frac{d}{dt}\, \left(
                        \begin{array}{c}
                          A_1(t) \\
                          A_2(t) \\
                        \end{array}
                      \right) =  {H}_{eff}(t)\,\left(
                        \begin{array}{c}
                          A_1(t) \\
                          A_2(t) \\
                        \end{array}
                      \right) \,,\label{theff}\ee where the matrix elements of $ {H}_{eff}(t)$ are obtained from equations (\ref{amp1},\ref{amp2}). Unlike the
                      case of a single species analyzed in detail in appendix (\ref{app:single}), for two   species mixing, $H_{eff}(t)$ is a $2 \times 2$ matrix, and $[ {H}_{eff}(t), {H}_{eff}(t')]\neq 0$ for $t\neq t'$, therefore the solution of the evolution equations is not a simple exponential. The usual approach, following the main approximation in the Weisskopf-Wigner method implemented in the (LOY) formulation\cite{lee}, invokes the
 long time limit\footnote{This   approximation is also implicitly implemented in ref.\cite{lee}.}
 \be \int^{t}_0    e^{i(E_b-E_{\kappa})(t-t' )}\,dt'~{}_{\overrightarrow{t\rightarrow \infty }}~ i\,\Bigg[\mathcal{P}\,\Big(\frac{1}{E_b-E_\kappa} \Big) -i \pi \delta(E_b-E_\kappa) \Bigg]\,,\label{pp} \ee yielding
 \be -i\,W_{ab}[t;t] \rightarrow \Delta_{ab}(E_b) \,, \label{Ws} \ee where $\Delta_{ab}(\omega) $ is defined by eqn. (\ref{sigofnu}).
   Taking this long time limit, the amplitude equations (\ref{theff}) become  an effective Schroedinger equation with a time \emph{independent} effective Hamiltonian
\be i\,\frac{d}{dt}\, \Bigg( \begin{array}{c}
                         A_1(t) \\
                         A_2 (t)
                      \end{array}\Bigg) = \mathcal{H}_{eff} \, \Bigg( \begin{array}{c}
                         A_1(t) \\
                         A_2 (t)
                      \end{array}\Bigg) \label{shcamp}\ee with

\be  \mathcal{H}_{eff} = \left(
                           \begin{array}{cc}
                             E_1+\Delta_{11}(E_1)  & \Delta_{12}(E_2) \\
                            \Delta_{21}(E_1)  & E_2+\Delta_{22}(E_2)  \\
                           \end{array}
                         \right) \equiv \left(
                                          \begin{array}{cc}
                                            \mathcal{H}_{11} & \mathcal{H}_{12} \\
                                            \mathcal{H}_{21} & \mathcal{H}_{22} \\
                                          \end{array}
                                        \right) = H_{eff}(\infty)\,.
  \label{Heffe}\ee
This effective Hamiltonian is not Hermitian, this is a manifestation that it describes the (approximate) dynamics of a \emph{quantum open system}, namely of a subset of degrees of freedom which are coupled to a continuum of other degrees of freedom whose dynamics has been ``integrated out''. Time evolution is not unitary in this subset, as is explicit from the unitarity condition (\ref{totder},\ref{unitarity}), which indicates a flow of probability from the $\ket{\phi_1},\ket{\phi_2}$ to the excited intermediate states $\ket{\big\{\kappa\big\}}$ which have been integrated out in the equations of motion.

  It proves convenient to rewrite $\mathcal{H}_{eff}$ as
  \be \mathcal{H}_{eff} = \frac{1}{2}\,\Big(E_1+\Delta_{11}(E_1)+E_2+\Delta_{22}(E_2) \Big)\,\mathbb{I}+ \frac{1}{2}\,\wD(E_1,E_2)\, \wR(E_1,E_2)\,,\label{HR}
   \ee where $\mathbb{I}$ is the $2\times 2$ identity matrix and
  \be \wD(E_1,E_2) = \Bigg[\Big(E_1+\Delta_{11}(E_1)-E_2-\Delta_{22}(E_2) \Big)^2+ 4\,\Delta_{12}(E_2)\Delta_{21}(E_1) \Bigg]^{1/2}\,,\label{wiD}  \ee and
   \be \wR(E_1,E_2)  = \left(
                                                                                                                         \begin{array}{cc}
                                                                                                                           \walpha(E_1,E_2) & \wbeta(E_1,E_2) \\
                                                                                                                           \wgamma(E_1,E_2) & -\walpha(E_1,E_2) \\
                                                                                                                         \end{array}
                                                                                                                       \right)\,,\label{wiR}\ee with the definitions
\bea \walpha(E_1,E_2) & = & \frac{\Big(E_1+\Delta_{11}(E_1)-E_2-\Delta_{22}(E_2) \Big)}{\wD(E_1,E_2)}\,\label{walfa}\\
\wbeta(E_1,E_2) & = & \frac{2\,\Delta_{12}(E_2)}{\wD(E_1,E_2)}~~; ~~ \wgamma(E_1,E_2)   =   \frac{2\,\Delta_{21}(E_1)}{\wD(E_1,E_2)}\,.\label{wbg}
\eea

It follows from these definitions that
\be \walpha^2+\wbeta\,\wgamma =1\,,\label{wrela} \ee which implies that
\be \wR^2(E_1,E_2) = \mathbb{I} \,,\label{wr2}\ee therefore the matrix $\wR$ features eigenvalues $\pm 1$.

  Consider the eigenvalue equation (suppressing the arguments $E_{1,2}$),
  \be  \wR\, \left( \begin{array}{c}
                                                    p^\pm \\
                                                    \pm\,q^\pm
                                                  \end{array}\right) = \pm  \,  \left( \begin{array}{c}
                                                    p^\pm \\
                                                    \pm \, q^\pm
                                                  \end{array}\right)\,,
                                         \label{eigen}\ee
 the solution of which is
 \bea && p^+ = N^+ \,(1+\walpha)~~;~~ q^+ = N^+\, \wgamma \,,\label{pqplu}\\&& p^- = N^-\,(1-\walpha)~~;~~ q^- = N^-\, \wgamma \,,\label{pqmin}\eea with $N^{\pm}$ normalization factors. These   are eigenvectors of $\mathcal{H}_{eff}$, namely
 \be \mathcal{H}_{eff} \, \left( \begin{array}{c}
                                                    p^\pm \\
                                                    \pm \, q^\pm
                                                  \end{array}\right) = \lambda^\pm\, \left( \begin{array}{c}
                                                    p^\pm \\
                                                    \pm \, q^\pm
                                                  \end{array}\right) \,,\label{Heigen}\ee with eigenvalues

 \be \lambda^\pm = \frac{1}{2}\,\Big[\Big(E_1+\Delta_{11}(E_1)+E_2+\Delta_{22}(E_2) \Big)\, \pm\,\wD(E_1,E_2) \Big] \equiv \widetilde{\varepsilon}^{\,\pm}-i\, \frac{\widetilde{\Gamma}^{\,\pm}}{2}\,,\label{lambdas}\ee where $\widetilde{\varepsilon}^{\,\pm};\widetilde{\Gamma}^{\,\pm}$ are both real. The effective Hamiltonian can be diagonalized by introducing
 \be U^{-1} = \left(
                \begin{array}{cc}
                   p^+  &  p^- \\
                   q^+  & -q^-  \\
                \end{array}
              \right)~~;~~ U =  \frac{1}{p^+\,q^- + q^+ \,p^- }\, \left(
                \begin{array}{cc}
                    q^- &   p^-  \\
                    q^+   & -p^+   \\
                \end{array}
              \right) \,,\label{Us}\ee satisfying $UU^{-1}=U^{-1}U = \mathbb{I}$, yielding
              \be U \mathcal{H}_{eff}U^{-1} =  \left(
                \begin{array}{cc}
                   \lambda^+ &  0 \\
                   0  & \lambda^- \\
                \end{array}
              \right)\,.\label{hdiag}\ee

 Let us define
 \be \Bigg( \begin{array}{c}
                         A_1(t) \\
                         A_2 (t)
                      \end{array}\Bigg) = U^{-1}\,\Bigg( \begin{array}{c}
                        V_1(t) \\
                        V_2 (t)
                      \end{array}\Bigg)\,, \label{AVrela} \ee the effective evolution equations for $V_{1,2}(t)$     become

 \be i\,\frac{d}{dt}\, \Bigg( \begin{array}{c}
                        V_1(t) \\
                        V_2 (t)
                      \end{array}\Bigg) =   \, \Bigg( \begin{array}{c}
                        \lambda^+\, V_1(t) \\
                       \lambda^-\, V_2(t)
                      \end{array}\Bigg) \Rightarrow  \Bigg( \begin{array}{c}
                        V_1(t) \\
                        V_2 (t)
                      \end{array}\Bigg) =  \left(
                                             \begin{array}{cc}
                                               e^{-i\lambda^+ t} & 0 \\
                                              0 & e^{-i\lambda^- t} \\
                                             \end{array}
                                           \right)
                         \Bigg( \begin{array}{c}
                        V_1(0) \\
                        V_2 (0)
                      \end{array}\Bigg)  \,. \label{shcampV}\ee  Using the   definition (\ref{AVrela}) evaluated at $t=0$  yields the solution for the amplitudes
  \be
  \Bigg( \begin{array}{c}
                       A_1(t) \\
                        A_2(t)
                      \end{array}\Bigg) =   U^{-1}\, \left(
                                             \begin{array}{cc}
                                               e^{-i\lambda^+ t} & 0 \\
                                              0 & e^{-i\lambda^- t} \\
                                             \end{array}
                                           \right) \,U \, \Bigg( \begin{array}{c}
                         A_1(0) \\
                         A_2(0)
                      \end{array}\Bigg)\,.
 \ee With the relations (\ref{wrela},\ref{pqplu},\ref{pqmin}) it is straightforward to find that
  \be  \Bigg( \begin{array}{c}
                       A_1(t) \\
                        A_2(t)
                      \end{array}\Bigg) =  \Big[e^{-i\lambda^+ t}\,\mathbb{\wP}_+ + e^{-i\lambda^- t}\,\mathbb{\wP}_- \Big] \Bigg( \begin{array}{c}
                         A_1(0) \\
                         A_2(0)
                      \end{array}\Bigg)\,,\label{asfina} \ee  with the projector operators
                      \be \mathbb{\wP}_{\pm} = \frac{1}{2}(\mathbb{I}\pm \wR)~~;~~ \mathbb{\wP}^{\,2}_{\pm} = \mathbb{\wP}_\pm\,,\label{wprojs}\ee where $\wR$ is given by eqn. (\ref{wiR}), or, alternatively

 \be  \Bigg( \begin{array}{c}
                       A_1(t) \\
                        A_2(t)
                      \end{array}\Bigg) =  \frac{1}{2}\Big[\Big(e^{-i\lambda^+ t}+e^{-i\lambda^- t} \Big) +\Big(e^{-i\lambda^+ t}-e^{-i\lambda^- t} \Big) \, \,\mathbb{\wR} \Big] \Bigg( \begin{array}{c}
                         A_1(0) \\
                         A_2(0)
                      \end{array}\Bigg)\,. \label{asfinalt} \ee
                      Comparing the results via Laplace transform and Breit-Wigner approximation, namely (\ref{solLT}) to the solution of the set of equations (\ref{shcamp}) obtained in the infinite time limit, namely (\ref{asfina},\ref{asfina}) we find several sources of discrepancies:

                      \vspace{1mm}

                      \textbf{i:)} the wavefunction  renormalization constants $Z_{\pm}$ in (\ref{solLT},\ref{solLTnd}) are missing in (\ref{asfina},\ref{asfinalt}).

\vspace{1mm}

                       \textbf{ii:)} Whereas the projector operators in (\ref{solLT}) depend on the values of $\omega_{\pm}$, namely the complex poles, those in (\ref{asfina}) depend on $E_1,E_2$ \emph{separately}. Furthermore, the values of the complex frequencies $\omega_{\pm}$ (\ref{bwpoles}) are not obviously similar to $\lambda^\pm$ (\ref{lambdas}).

                      The origin of these discrepancies can be traced to taking the long time limit (\ref{pp},\ref{Ws}) \emph{before} integrating the set of equations (\ref{amp1},\ref{amp2}), which are equivalent to the original set of equations (\ref{dotC1},\ref{dotC2}) up to order $\mathcal{O}(H^2_I)$.  Any discrepancy between the order of the long-time limits will translate into differences of $\mathcal{O}(H^2_I)$.

                      In appendix (\ref{app:single}) it is shown that the discrepancy  in wave function renormalization of amplitudes originates in this long time limit in the simpler case of one species. We now compare the results for the eigenvalues and eigenvectors of the Laplace transform method and the Markov approximation with the effective Hamiltonian.

\vspace{1mm}

\textbf{Non-degenerate case:} For $E_1-E_2 \gg \Delta_{ab}$ we can approximate $\widetilde{D}(E_1,E_2)$, given by (\ref{wiD}) as
\be \widetilde{D}(E_1,E_2) \simeq E_1+\Delta_{11}(E_1)-E_2-\Delta_{22}(E_2)\,, \label{ditil} \ee    yielding for the eigenvalues $\lambda^\pm$, eqn. (\ref{lambdas})
\be \lambda^+ =  E_1+\Delta_{11}(E_1)  \equiv E^R_1-i \frac{\Gamma^+}{2}    ~~;~~ \lambda^- = E_2+\Delta_{22}(E_2) \equiv E^R_2-i \frac{\Gamma^-}{2} \,,\label{lamnonde}\ee   which agree with the eigenvalues obtained from the Laplace transform   (\ref{wpnodeg},\ref{wmnodeg}). For the amplitudes we now find up to $\mathcal{O}(\Delta)$
 \bea A_1(t) & = &  \,    \,A_1(0)\,e^{-i\lambda_+ t} +\frac{1}{2}\,\Bigg[  \frac{\Delta_{12}(E_2)\,\,e^{-i\lambda_+ t}-\Delta_{12}(E_2)\,\,e^{-i\lambda_- t}}{E^R_1-E^R_2} \Bigg]\,A_2(0)  \,,\label{A1texmkv}\\ A_2(t) & = &      \,A_2(0)\,e^{-i\lambda_- t} +\frac{1}{2}\,\Bigg[  \frac{\Delta_{21}(E_1)\,\,e^{-i\lambda_+ t}-\Delta_{21}(E_1)\,\,e^{-i\lambda_- t}}{E^R_1-E^R_2} \Bigg]\,\,A_1(0)   \,.\label{A2texmkv}
 \eea The differences with the result from the Laplace transform, equations (\ref{A1tex},\ref{A2tex}) are noteworthy: \textbf{i:}) the wave function renormalization constants multiplying the diagonal terms in (\ref{A1tex},\ref{A2tex}) are missing in (\ref{A1texmkv},\ref{A2texmkv}), \textbf{ii:)}  the differences in the arguments of $\Delta_{12},\Delta_{21}$ in the brackets. Clearly the discrepancies are of second order in $\Delta_{ab} \propto H^2_I$, as discussed above.

 \vspace{1mm}

 \textbf{(Nearly) degenerate case:} For $E_1,E_2 \gg \Delta_{ab}$ but with $E_1-E_2 \equiv \delta \lesssim \mathcal{O}(\Delta)$, it follows that
 \be \lambda^\pm = \overline{E}+ \frac{1}{2}\Big(\Delta_{11}(\overline{E})+\Delta_{2}(\overline{E}) \Big)\pm \frac{\widetilde{D}(\overline{E})}{2}~;~ \overline{E} = \frac{1}{2}(E_1+E_2)\,,\label{lampmdeg}\ee where $\widetilde{D}(\overline{E})$ corresponds to setting $E_1\simeq E_2 \simeq \overline{E}$ in the matrix elements of $\widetilde{D}(E_1,E_2)$. The eigenvalues $\lambda^\pm$ again coincide with $\omega^\pm$ given by eqn. (\ref{polesdeg}). Furthermore, it is straightforward to confirm that in this case $\wD(E_1,E_2) = \mathbb{R}(\overline{E})$ given by eqn. (\ref{capRE}). Therefore, the main difference between the Laplace result (\ref{solLTnd}) and that from the effective Hamiltonian (\ref{asfinalt}) is the wave function renormalization $Z$ multiplying the initial amplitudes in (\ref{solLTnd}).

 In fact, at a fundamental level, the emergence of the wave function renormalization of the amplitudes of the quasinormal modes precludes    the description of their time evolution in terms of an effective non-Hermitian Hamiltonian. This can be understood from the following simple argument: the formal solution of the amplitude equation (\ref{shcamp}) is
 \be \Bigg( \begin{array}{c}
                         A_1(t) \\
                         A_2 (t)
                      \end{array}\Bigg) =  e^{-i\mathcal{H}_{eff}\,t}\,  \Bigg( \begin{array}{c}
                         A_1(0) \\
                         A_2 (0)
                      \end{array}\Bigg) \,,\label{soluA}  \ee  which obviously does not include a wave function renormalization as pre-factor of the quasinormal mode amplitudes. The wave function renormalization is an off-shell contribution that describes the dressing by virtual states of the single (quasi) particles on short time scales, and yields second order corrections to the amplitudes. While it may be finite in the case of the box diagram contribution to flavored neutral meson mixing, it is in general ultraviolet divergent in quantum field theory.


                     Therefore we conclude that the Laplace transform with the Breit-Wigner approximation provides a more accurate description of the evolution of mixing as compared to that obtained from the effective non-Hermitian Hamiltonian.

\vspace{1mm}

\textbf{Quantum beats:} The two orthogonal states $\ket{\phi_1},\ket{\phi_2}$ decaying into a common channel $\ket{\kappa}$ lead to interference in the amplitudes of the decay state $\ket{\kappa}$ as a consequence of  ``which path'' information in the decay. This is similar to the case of quantum beats in  ``V''-shaped three level systems, in which two higher levels radiatively decay to the lowest level\cite{zubairy}, an ubiquitous phenomenon in quantum optics. This interference phenomenon, or quantum beats, is  featured in the amplitudes of the decay products described by the states, $\ket{\kappa}$, namely the coefficients $C_{\kappa}(t)$.

The analysis above has focused on the time evolution of the amplitudes $C_{1,2}(t)$, which also determine the amplitudes $C_{\kappa}(t)$ of the intermediate states via equation (\ref{cmoft}). Writing these coefficients in terms of the amplitudes $A_{1,2}(t)$, and introducing the spectral densities (\ref{rhosd}) we find
\be \sum_{\kappa} |C_{\kappa}(t)|^2 = \int^t_{0}dt_1 \int^t_0 dt_2 \int^{\infty}_{-\infty} dk_0\, \,\sum_{a,b=1,2}\,A^*_a(t_1)\,\mu_{ab}(k_0)\,A_b(t_2)\,e^{-ik_0(t_1-t_2)}\,. \label{kstatprob}\ee It is convenient to introduce $\Theta(t_1-t_2)+\Theta(t_2-t_1) =1$ inside the time integrations,  use the property of the spectral density (\ref{rhosd}),  and the definition of the self-energy (\ref{sigmasd}) to show that
\be \sum_{\kappa} |C_{\kappa}(t)|^2 = \sum_{a,b=1,2}\int^t_{0}dt_1   A^*_a(t_1)\,\int^{t_1}_0 \sigma_{ab}(t_1-t_2)\,A_b(t_2)\,dt_2 + c.c.\,,\label{sumnu}\ee the complex conjugate (c.c) contribution arises from the term with $\Theta(t_2-t_1)$ upon relabelling $t_1 \leftrightarrow t_2$, $a\leftrightarrow b$ and using the property (\ref{rhosd}).  Using the amplitude equations (\ref{equafin}) we finally find,
\be \sum_{\kappa} |C_{\kappa}(t)|^2 = -\sum_{a=1,2}\int^t_{0}dt_1 \frac{d}{dt_1}\Big[A^*_a(t_1)\,A_a(t_1) \Big] = |A_1(0)|^2+|A_2(0)|^2 - \Big[|A_1(t)|^2+|A_2(t)|^2\Big]\,,  \label{intunit}\ee this result is precisely  the unitarity relation formally established by eqns. (\ref{totder},\ref{unitarity}) providing a complementary and explicit proof of unitarity exhibiting the role of the self-energy.

Following (LOY)\cite{lee}, introducing the total population  of the $\phi_1,\phi_2$ states as
\be  {N}(t) = \Big[|A_1(t)|^2+|A_2(t)|^2\Big]\equiv \Big[|C_1(t)|^2+|C_2(t)|^2\Big]\,, \label{ocuK}\ee   and writing  the amplitudes $A_a(t)$ as linear superpositions of the quasinormal modes, namely
\be A_a(t) = A_{a+}\,e^{-i\,\varepsilon_{+} t}\,e^{-\frac{\Gamma_+ }{2}t}+A_{a-}\,e^{-i\,\varepsilon_{-} t}\,e^{-\frac{\Gamma_- }{2}t}~;~ a=1,2\,,\label{combo}\ee
where the coefficients $A_{a\pm}$ can be read off eqn. (\ref{solLT}), it follows that
\be N(t) = \sum_{a=1,2}\Big[|A_{a+}|^2\,e^{-\Gamma_+ t} + |A_{a-}|^2\,e^{-\Gamma_- t}+2\,\mathrm{Re}\Big(A^*_{a+}A_{a-}\,e^{i(\varepsilon_+-\varepsilon_-)\,t} \Big)\,e^{-(\Gamma_++\Gamma_-)\,t/2}  \Big]\,,\label{populee}\ee the last term displays the \emph{quantum beats} as a consequence of the interference between the quasinormal modes.
 With the normalization (\ref{unit}) the unitarity relations (\ref{intunit},\ref{unitarity}) yield
\be  \sum_{\kappa} |C_{\kappa}(t)|^2 = 1 - N(t)\,, \label{qbits}\ee  displaying the quantum beats from (\ref{populee}) in the last term. Therefore unitarity entails that the quantum beats in the total population are reflected in the time evolution of the decay products.

These interference terms are, of course, well known, originally recognized in the seminal work by (LOY)\cite{lee}  and have been experimentally observed in the decays products of flavored neutral mesons\cite{cprev,cpv1,cpv2,pdg}. We note that the coefficients $A_{a\pm}$ depend on the wave function renormalization constants $Z_{\pm}$ in the solutions (\ref{solLT}), an important discrepancy with the usual effective non-hermitian Hamiltonian description of particle mixing.

Our main objective in analyzing the dynamics of mixing within the framework of the (LOY) theory of flavored meson mixing,  is to provide a guide to and benchmark for the effective field theory approach to the  dynamics   of mixing in a medium studied in the next section.

\section{The   Effective Action for particle mixing.}\label{sec:effaction}
The previous section extended and generalized the formulation of particle mixing, originally implemented to study CP violation in the neutral Kaon system, to the
case in which different particles (in general with different masses) mix via   common intermediate states or decay channels. As it is clear from this analysis, such formulation is applicable and generally applied to the case of an initial state being a pure state, and primarily, when such state is a   linear superposition of single particle states\cite{lee}. This analysis also revealed several subtleties associated with the time evolution of the amplitudes in terms of an effective Hamiltonian. It also highlighted that the non-hermiticity of the effective Hamiltonian is a hallmark of a \emph{quantum open system}, namely such Hamiltonian describes the non-unitary time evolution of a reduced subset of states which are coupled to a continuum of states that have been ``integrated out''.

Our main objective is to provide a framework to study the dynamics of particle mixing in a medium, as it is necessary within the realm of cosmology. In this case, we are interested in the time evolution of a density matrix, describing a statistical ensemble of particles, not just a pure state of few particles. Furthermore, we are interested in obtaining the time evolution of correlation functions  and distribution functions   in the medium, in particular their asymptotic behavior and possible thermalization, not on the amplitudes of single (or few) particle states.

Rather than considering the most general case of mixing between charged bosons or fermions which necessarily add several technical complications, we consider the simpler case of  real  scalar or pseudoscalar bosonic fields  $\phi_1,\phi_2$ interacting with   degrees of freedom in thermal equilibrium denoted collectively by   $\chi$, to establish the main framework and results within a simpler setting, thus paving the way to extrapolating to a more general case.

The mixing between $\phi_1$ and $\phi_2$ is indirect and a consequence of  a coupling to a common set  of intermediate states yielding a self-energy with off-diagonal elements in the space of $\phi_{1,2}$ similar to the cases studied in the previous section.

The  general Lagrangian density describing this situation is given by
\begin{equation}
    \mathcal{L}[\phi_1,\phi_2,\chi] = \mathcal{L}_\phi + \mathcal{L}_{\chi} + \mathcal{L}_I \,,\label{lag}
\end{equation}
where
\bea
    \mathcal{L}_\phi
    & = & \frac{1}{2}\,\sum_{a=1,2}     \, \Big[ \big(\partial \phi_a\big)^2 -   m^2_a\, \phi^2_a  \Big]   \nonumber    \\
    \mathcal{L}_I
    & = &  - \phi_1 \mathcal{O}_{1}[\chi] - \phi_2 \mathcal{O}_{2}[\chi]\,,\label{lagra}
\eea
 where  $\mathcal{L}_\chi$ is the Lagrangian of the $\chi$ fields. These are assumed to describe  degrees of freedom in thermal equilibrium including interactions among these fields, and  $\mathcal{O}_{1,2}[\chi] $ are (composite) operators associated with the $\chi$ degrees of freedom. These operators include couplings $g_{1,2}$ assumed to be small.  Indirect field mixing is a consequence of non-vanishing correlations $\langle \mathcal{O}_1\,\mathcal{O}_2 \rangle$ in the medium yielding off diagonal self-energy matrix elements.

  Let us consider the initial density matrix at a time
$t=0$ to be of the form
\begin{equation}
\hat{\rho}(0) = \hat{\rho}_{\phi}(0) \otimes
\hat{\rho}_{\chi}(0) \,.\label{inidensmtx}
\end{equation}

The initial density matrix $\hat{\rho}_\phi(0)$ is normalized so that  $\mathrm{Tr}_\phi \hat{\rho}_\phi(0) =1$ and that of the $\chi$ fields will be taken to
describe a statistical ensemble in thermal equilibrium at a temperature
$T=1/\beta$, namely

\begin{equation}
\hat{\rho}_{\chi}(0) = \frac{e^{-\beta\,H_{\chi}}}{\mathrm{Tr}_{\chi} e^{-\beta H_\chi}}\,,\label{rhochi}
\end{equation}

\noindent where $H_{\chi} $ is the total Hamiltonian for the
fields $\chi$, and may include other fields to which $\chi$ is coupled other than the  fields $\phi_{1,2}$. The $\chi$-vacuum is obtained in the limit $\beta \rightarrow \infty$.

 For example, for the discussion of the previous section the initial density matrix is given by
 \be \hat{\rho}(0) = \ket{\Psi(t=0)}\bra{\Psi(t=0)} \,, \label{vacro}\ee where $\ket{\Psi(t=0)}$ is the state (\ref{inistate}).

The factorization of the initial density matrix is an assumption often explicitly or implicitly made in the literature, it can be relaxed by including initial correlations among the various fields   at the expense of daunting technical complications. In this study we will not consider   this important case, assuming the factorization as in (\ref{rhochi}). In what follows we will refer to the set of fields $\phi_{1,2}$ collectively simply as $\phi\equiv \{\phi_1,\phi_2\}$ to simplify notation.

The main concept that anchors the framework developed below is the following: the time evolution of the full density matrix in the Schroedinger picture is given by
\be \hat{\rho}(t)= e^{-iHt}\,\hat{\rho}(0)\,e^{iHt}\,, \label{rhoft} \ee   where $H$ is the total Hamiltonian,
 \begin{equation}
H=H_{0 \phi} + H_{\chi}+ \int d^3x \sum_{a=1,2} \phi_{a} \mathcal{O}_{a}(\chi)  \,, \label{hami}
\end{equation}
where $H_{0 \phi},H_{\chi}$  are the   Hamiltonians for   the respective fields.   We will assume that the composite operators $\mathcal{O}_a$ include weak couplings so as to define a perturbative expansion, second order terms in $\mathcal{O}_a$ imply second order in couplings, which we will denote as $\mathcal{O}(g^2)$ with $g$ a generic coupling.

 The reduced density matrix for the $\phi_{1,2}$ degrees of freedom is obtained by tracing over the $\chi$ degrees of freedom, namely
\be \hat{\rho}^r_{\phi}(t) = \mathrm{Tr}_{\chi}\hat{\rho}(t) \,.\label{rhored}\ee This reduced density matrix does not evolve unitarily in time, its time evolution is determined by a time non-local  effective action\cite{beilok,schwinger,keldysh,maha,feyver,mario}. One of our main objectives is to obtain this effective action.

  It is  convenient to write the density matrix in the  field basis which     facilitates a path integral representation of the non-equilibrium reduced density matrix\cite{beilok,schwinger,keldysh,maha,feyver,mario}.

  In the field basis the matrix elements of $\hat{\rho}_{\phi}(0)$ and $\hat{\rho}_{\chi}(0)$
are given by
\begin{equation}
\langle \phi |\hat{\rho}_{\phi}(0) | \phi'\rangle =
\rho_{\phi,0}(\phi ,\phi')~~;~~\langle \chi|\hat{\rho}_{\chi}(0) | \chi'\rangle =
\rho_{\chi,0}(\chi ;\chi')\,,
\end{equation}   this is a \emph{functional} density matrix as the fields feature spatial arguments.
$\hat{\rho}_\phi(0)$    represents either  a pure state, such as coherent state,  or more generally an initial
 statistical ensemble, whereas $\hat{\rho}_{\chi}(0)$ is assumed to describe a thermal ensemble and is given by eqn. (\ref{rhochi}).

  To obtain the effective action,   we  follow the procedure described above:  evolve the initial density matrix in time,  trace over the $\chi$ degrees of freedom thereby obtaining the reduced density matrix for the $\phi$  fields, and   the effective action from its time evolution. Including source terms for the fields $\phi$,
 we can compute   expectation values and correlation functions as a function of time from variational derivatives as usual.

 We now follow the main methods and results of refs.\cite{shuyang,mesonmix}, summarizing here the main aspects pertinent to the case of mixing for consistency of presentation.

The  {reduced} density matrix is given by
\be \rho^{r}_{\phi}(t) = \mathrm{Tr}_{\chi} U(t)\, \hat{\rho}(0)\,U^{-1}(t)~~;~~ U(t)= e^{-iHt}\,.  \label{rored}\ee In field space,

\be    \rho(\phi_f,\chi_f;\phi'_f,\chi'_f;t)   =        \langle \phi_f;\chi_f|U(t)\hat{\rho}(0)U^{-1}(t)|\phi'_f;\chi'_f\rangle \,,\label{evolrho} \ee from which the reduced density matrix elements are obtained by taking the trace on $\chi$, namely setting $\chi'_f = \chi_f$ and carrying out the functional integral in $\chi_f$,
\be \rho^r(\phi_f ;\phi'_f,;t) = \int D\chi_f \,\langle \phi_f;\chi_f|U(t)\hat{\rho}(0)U^{-1}(t)|\phi'_f;\chi_f\rangle \,. \label{redmtxel}\ee

With the functional integral representation
\bea \langle \phi_f;\chi_f|U(t)\hat{\rho}(0)U^{-1}(t)|\phi'_f;\chi'_f\rangle  & = &  \int D\phi_i D\chi_i D\phi'_i D\chi'_i ~ \langle \phi_f;\chi_f|U(t)|\phi_i;\chi_i\rangle\,\rho_{\phi,0}(\phi_i;\phi'_i)\, \otimes   \nonumber \\ & & \rho_{\chi,0}(\chi_i;\chi'_i) \,
 \langle \phi'_i;\chi'_i|U^{-1}(t)|\phi'_f;\chi'_f\rangle \,,\label{evolrhot}\eea it follows that the reduced
 density matrix elements are
\bea \rho^r(\phi_f ;\phi'_f,;t) & = & \int D\chi_f   \int D\phi_i D\chi_i D\phi'_i D\chi'_i ~ \langle \phi_f;\chi_f|U(t)|\phi_i;\chi_i\rangle\,\rho_{\phi,0}(\phi_i;\phi'_i)\, \otimes   \nonumber \\ & & \rho_{\chi,0}(\chi_i;\chi'_i) \,
 \langle \phi'_i;\chi'_i|U^{-1}(t)|\phi'_f;\chi_f\rangle \,.\label{redfun} \eea

  The $\int D\phi$ etc, are functional integrals where the spatial argument has been suppressed. The matrix elements of the time evolution forward and backward can be written as path integrals, namely
 \bea   \langle \phi_f;\chi_f|U(t)|\phi_i;\chi_i\rangle  & = &    \int \mathcal{D}\phi^+ \mathcal{D}\chi^+\, e^{i \int d^4 x \mathcal{L}[\phi^+,\chi^+]}\label{piforward}\\
 \langle \phi'_i;\chi'_i|U^{-1}(t)|\phi'_f;\chi'_f\rangle &  =  &   \int \mathcal{D}\phi^- \mathcal{D}\chi^-\, e^{-i \int d^3 x \mathcal{L}[\phi^-,\chi^-]}\label{piback}
 \eea where we use the shorthand notation
 \be \int d^4 x \equiv \int_0^t dt \int d^3 x \,,\label{d4xdef}\ee
 $ \mathcal{L}[\phi,\chi] $ is given by (\ref{lag},\ref{lagra})   and
 the boundary conditions on the path integrals are
  \bea     \phi^+(\vec{x},t=0)=\phi_i(\vec{x})~;~
 \phi^+(\vec{x},t)  &  =  &   \phi_f(\vec{x})\,,\label{piforwardbc}\\   \chi^+(\vec{x},t=0)=\chi_i(\vec{x})~;~
 \chi^+(\vec{x},t) & = & \chi_f(\vec{x}) \,,\label{chipfbc}  \\
     \phi^-(\vec{x},t=0)=\phi'_i(\vec{x})~;~
 \phi^-(\vec{x},t) &  = &    \phi'_f(\vec{x})\,,\label{aminbc} \\   \chi^-(\vec{x},t=0)=\chi'_i(\vec{x})~;~
 \chi^-(\vec{x},t) & = & \chi'_f(\vec{x}) \,.\label{pibackbc}
 \eea

The field variables $\phi^\pm, \chi\pm$ along the forward ($+$) and backward ($-$) evolution branches are recognized as those necessary for the in-in or  Schwinger-Keldysh\cite{schwinger,keldysh,maha,beilok} closed time path approach to the time evolution of a density matrix.

 The reduced density matrix for the  fields $\phi_a$ (\ref{redfun}),    can be written as
 \be   \rho^{r}(\phi_f,\phi'_f;t) = \int D\phi_i   D\phi'_i  \,  \mathcal{T}[\phi_f,\phi'_f;\phi_i,\phi'_i;t] \,\rho_\phi(\phi_i,\phi'_i;0)\,, \label{timevrho} \ee
where the time evolution kernel is given by
\be \mathcal{T}[\phi_f,\phi_i;\phi'_f,\phi'_i;t] = {\int} \mathcal{D}\phi^+ \, \int \mathcal{D}\phi^- \, e^{i  S_{eff}[\phi^+,\phi^-;t]}\,, \label{tauop} \ee
from which the \emph{in-in effective action} out of equilibrium is identified as
\begin{equation}
    {S}_{eff}[\phi^+,\phi^-;t] = \int^t_0 dt' \int d^3 x \Big\{ \mathcal{L}_0[\phi^+]-\mathcal{L}_0[\phi^-] +\mathcal{I}[ \phi^+, \phi^-;t] \Big\} \,,\label{Leff}
\end{equation} where
$\mathcal{I}[\phi^+;\phi^-;t]$ is the \emph{influence action}\cite{beilok,feyver} and  is obtained by tracing over the $\chi$ degrees of freedom, namely
\bea
    e^{i\mathcal{I}[\phi^+;\phi^-;t]} & = &  \int D\chi_i \,D\chi'_i D\chi_f  \int \mathcal{D}\chi^+ \int \mathcal{D}\chi^- \, e^{i  \int d^4x \left[\mathcal{L}[\chi^+]-  \sum_{a}\phi^+_a\,\mathcal{O}_a[\chi^+] \right]}\nonumber \\  & \times &  e^{-i  \int d^4x \left[\mathcal{L}[\chi^-]-\sum_{a}\phi^-_a\,\mathcal{O}_a[\chi^-] \right]} \,   \rho_{\chi}(\chi_i,\chi'_i;0) \,,
    \label{inffunc}
\eea with the definition (\ref{d4xdef}).

\begin{figure}[ht]
\includegraphics[height=3.0in,width=3.0in,keepaspectratio=true]{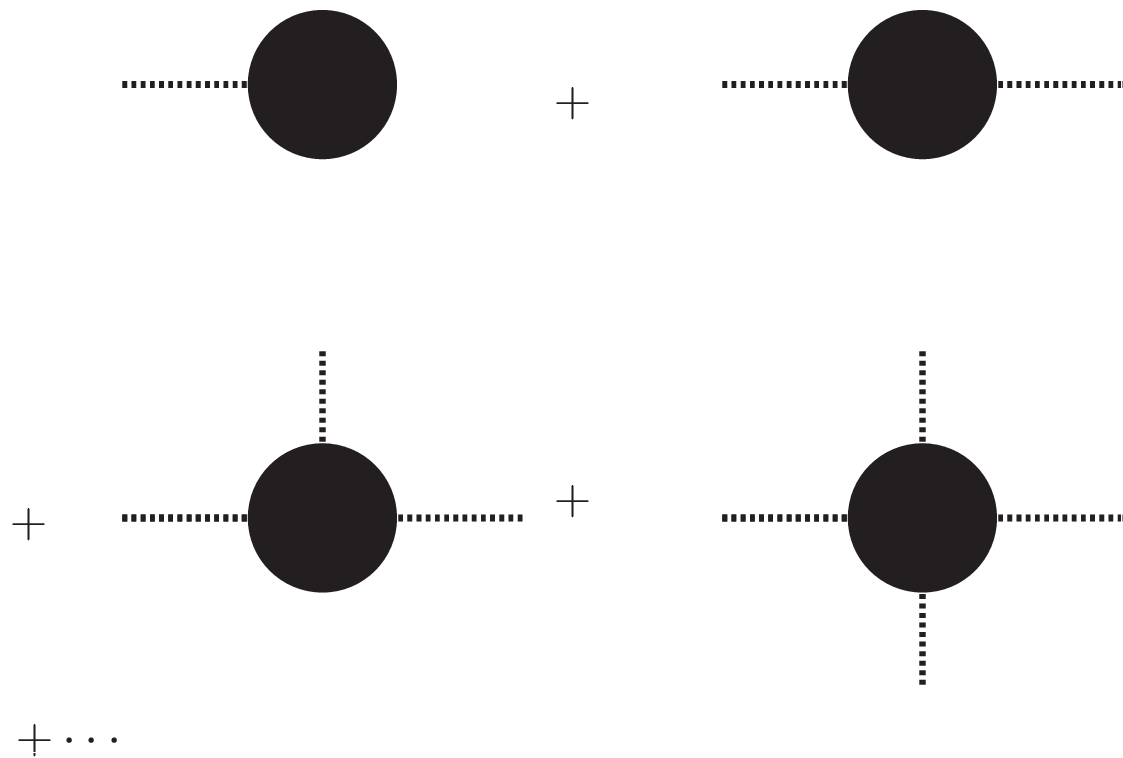}
\caption{Pictorial representation of the influence action $\mathcal{I}[\phi^+;\phi^-;t]$. The dashed lines are the (background) fields $\phi^\pm_a$, the filled circles the trace over the $\chi$ fields yielding correlation functions of the operators $\mathcal{O}[\chi^\pm]$. Each vertex carries a coupling. The second graph with two dashed lines yields the influence function up to second
order in  the couplings (\ref{Funravel}). We assume that $\langle \mathcal{O}_a\rangle =0$, hence there is no first order contribution to $\mathcal{I}[\phi^+;\phi^-;t]$.  }
\label{fig:effective}
\end{figure}

Note that the in the influence action (\ref{inffunc}) the $\phi$ fields act as \emph{background} field variables, the functional and path integrations are performed in all fields $\chi$  \emph{other} than
the $\phi$ fields. These functional integrals are obtained by expanding the terms $e^{\mp i \int d^4x \phi^\pm_a \mathcal{O}_a[\chi^\pm]}$ in power series and carrying the path and functional integrals in $\chi^\pm;\chi_i;\chi_f;\chi'$ yielding correlation functions of the operators $\mathcal{O}_a$ and to re-exponentiate. This is depicted in fig.(\ref{fig:effective})

 From equations (\ref{timevrho},\ref{tauop}) it is clear that the effective action $S_{eff}$ determines the time evolution of the reduced density matrix.

The path integral representations for both $\mathcal{T}[\phi_f,\phi_i;\phi'_f,\phi'_i;t]$ and $\mathcal{I}[\phi^+;\phi^-;t]$ feature the boundary conditions in (\ref{piforwardbc}-\ref{pibackbc}) except that we now set $\chi^\pm(\vec{x},t) = \chi_f(\vec{x})$ to trace over the $\chi$ field.

  The technical steps to obtain $\mathcal{I}[\phi^+;\phi^-;t]$ in perturbation theory, up to second order in the operators $\mathcal{O}_a$, up to $\mathcal{O}(g^2)$ are available in ref.\cite{mesonmix}. We follow the steps in this reference to find up to second order in couplings
\bea  && i\mathcal{I}[\phi^+, \phi^-;t]    =  -  \int d^4x_1 d^4x_2  \,\Bigg\{ \phi^+_a(\vx_1,t_1)\phi^+_b(\vx_2,t_2)\,G^>_{ab}(x_1-x_2) +   \phi^-_a(\vx_1,t_1)\phi^-_b(\vx_2,t_2)\,G^<_{ab}(x_1-x_2) \nonumber\\
  &- & \phi^+_a(\vx_1,t_1)\phi^-_b(\vx_2,t_2)\,G^<_{ab}(x_1-x_2)  -   \phi^-_a(\vx_1,t_1)\phi^+_b(\vx_2,t_2)\,G^>_{ab}(x_1-x_2)\Bigg\}\Theta(t_1-t_2) \label{Funravel}\eea where   and  $G^{\lessgtr}_{ab}(x_1-x_2)$ are given by
 \begin{eqnarray}
&&   G^>_{ab}(x_1-x_2) =  \langle
{\cal O}_{a}(x_1) {\cal O}_{b}(x_2)\rangle_\chi    \,,\label{ggreat} \\&&   G^<_{ab}(x_1-x_2)  =   \langle
{\cal O}_{b}(x_2) {\cal O}_{a}(x_1)\rangle_\chi     \,,\label{lesser}
\end{eqnarray}    and we have assumed that $\langle \mathcal{O}_a \rangle =0$ (so that the first diagram in fig. (\ref{fig:effective}) vanishes). The operators $\mathcal{O}_a$ are hermitian from which  it follows   that
\be G^<_{ab}(x_1-x_2) = G^>_{ba}(x_2-x_1)\,. \label{idt}\ee

   This is the general form of the influence function up to second order in the  operators $\mathcal{O}_a[\chi]$,    but \emph{to all orders} in the couplings of the $\chi$ fields  to \emph{any} other field except $\phi_{1,2}$.
 We can obtain expectation values and correlation functions of $\phi_{1,2}$  by including sources $J^{\pm}_a(x)$ with $\mathcal{L}_0(\phi^\pm)\rightarrow \mathcal{L}_0(\phi^\pm)\pm{\sum}_a J^{\pm}_a(x)\phi^\pm_a(x)$ and defining the generating functional
  \be \mathcal{Z}[J^+,J^-] = \mathrm{Tr}\,\rho^r(J^+,J^-;t) =  \int D\phi_f D\phi_i   D\phi'_i  \,  {\int} \mathcal{D}\phi^+ \, \int \mathcal{D}\phi^- \, e^{i S_{eff}[\phi^+,J^+;\phi^-,J^-;t]} \,\rho_\phi(\phi_i,\phi'_i;0) \label{Zofj}\ee with the boundary conditions
  \bea &  &   \phi^+_a(\vec{x},t=0)=\phi_{i,a}(\vec{x})~;~
 \phi^+_a(\vec{x},t)  =   \phi_{f,a}(\vec{x}) \nonumber \\
&  &   \phi^-_a(\vec{x},t=0)=\phi'_{i,a}(\vec{x})~;~
 \phi^-_a(\vec{x},t)  =   \phi_{f,a}(\vec{x}) \,.\label{bctraza}\eea Expectation values or correlation functions of $\phi^{\pm}$ in the reduced density matrix are obtained as usual with variational derivatives with respect to the sources $J^\pm$.

 The effective action (\ref{Leff}) may be written in a manner more suitable to exhibit the equations of motion by introducing the Keldysh\cite{keldysh} center of mass and relative variables
\be \Phi_a(\vx,t)= \frac{1}{2}\big( \phi^+_a(\vx,t)+\phi^-_a(\vx,t)\big)~~;~~\mathcal{R}_a(\vx,t)= \big( \phi_a^+(\vx,t)-\phi^-_a(\vx,t)\big)\label{kelvars} \,. \ee

The boundary conditions on the $\phi^\pm$ path integrals given by
(\ref{bctraza}) translate into the following boundary conditions on
the center of mass and relative variables
\begin{align}
    \Phi_a(\vec x,t=0)= \Phi_{a,i} \; \; &; \; \; \mathcal{R}_a(\vec x,t=0)=\R_{a,i} \,, \label{bcwig} \\
    \Phi_a(\vec{x},t=t_f) = \Phi_{a,f}(\vec{x}) \; \; &; \; \; \mathcal{R}_a(\vec x,t=t_f )=0 \,. \label{Rfin}
\end{align}

 Taking the spatial Fourier transform   the effective action (\ref{Leff})  with the influence functional (\ref{Funravel})   becomes
\bea
   && iS_{eff}[\Phi,\R] =
      - i \int d^3x \,\sum_{a}\,\mathcal{R}_{a,i}(x) \dot{\Phi}_a(x,t=0) \nonumber \\
    & & + i \int_0^t dt\, \sum_{\vec{k},a} \left\{ - \mathcal{R}_a(-\vec{k},t) \left( \ddot{\Phi}_{a}(\vec{k},t) + \omega^2_a(k) \Phi_{a}(\vec{k}, t)  \right)  +  \Phi_a (\vec{k},t) \J_a(-\vec{k},t) \right\} \nonumber \\
    && - \int_0^t dt_1 \int_0^t dt_2 \sum_{ab}\,\left\{ \frac{1}{2} \mathcal{R}_a(-\vec{k}, t_1) \mathcal{N}_{ab}(\vec{k};t_1-t_2) \R_b(\vec{k},t_2) + i\,\mathcal{R}_a(-\vec{k},t_1)   \Sigma^R_{ab}(\vec{k};t_1 - t_2) \Phi_b(\vec{k}; t_2) \right\}\,.\nonumber \\
    \label{efflanwig}
\eea where $\omega^2_{a}(k) = k^2+m^2_a$. To obtain the above form,  we integrated by parts in time,  defined $ \J_a(x)= (J^+_a(x)-J^-_a(x))$, and  kept only the sources conjugate to $\Phi_a$ because we are interested in expectation values and correlation functions of this variable only as discussed in detail below.

The non-local kernels in the above effective Lagrangian are given by \cite{mesonmix}
\begin{eqnarray}
\mathcal{N}_{ab}(k,t-t') & = &
\frac{1}{2} \left[   G^>_{ab}(k;t-t')+{ G}^<_{ab}(k;t-t') \right] \label{kernelkappa} \\
i\Sigma^{R}_{ab}(k,t-t') & = &    \left[{ G}^>_{ab}(k;t-t')-{
G}^<_{ab}(k;t-t') \right]\Theta(t-t') \equiv
i\Sigma_{ab}(k,t-t')\Theta(t-t') \label{kernelsigma}
\end{eqnarray} where $G^{<,>}(k;t-t')$ are the spatial Fourier transforms of the correlation functions in (\ref{ggreat},\ref{lesser}). It is clear from these correlation functions that if $\langle \mathcal{O}_1\,\mathcal{O}_2 \rangle \neq 0$ the self-energy features non-vanishing off diagonal matrix elements, these are responsible for  indirect mixing between the fields $\phi_1,\phi_2$. Since each operator is associated with a coupling $g_a$, the self-energy and noise kernels are of second order, and we will refer generically as $\Sigma \propto g^2;\mathcal{N}_{ab} \propto g^2$ to emphasize the second order nature of these kernels.

In the exponential of the effective action $e^{iS_{eff}}$, the quadratic term in the relative variables $\R_a$ can be written as a functional integral over a noise variable $\xi_a$ as follows,

\bea
    &&\exp\left\{ -  \frac{1}{2}\,\int dt_1 \int dt_2   \,\mathcal{R}_a(-\vec{k}; t_1) \mathcal{N}_{ab}(\vec{k};t_1 - t_2) \mathcal{R}_b(\vec{k};t_2) \right\} =     \nonumber \\
    &&   \widetilde{C} \int \mathcal{D}\xi_a \, \exp\left\{ - \frac{1}{2} \int dt_1 \int dt_2 \, \xi_a(-\vec{k};t_1) \mathcal{N}^{-1}_{ab}(\vec{k};t_1-t_2) \xi_b(\vec{k};t_2) + i \int dt \xi_a(-\vec{k};t) \R_a(\vec{k};t) \right\}\nonumber\\
    \label{nois}
\eea
where $\widetilde{C}$ is a normalization factor.

The time evolution of the density matrix defines an initial value problem, consequently  we seek to obtain the equations of motion as an initial value problem rather than a boundary value problem.  Since the Heisenberg equations of motion are second order in time, an initial value problem is determined by providing the initial values of the field and its canonical momentum. This suggests to consider the Wigner transform of the initial density matrix   by   writing  it in terms of the initial center of mass and relative variables $\Phi_{a,i},\R_{a,i}$
\be \rho_\phi(\phi_{a;i},\phi'_{a;i};0) \equiv \rho_\phi(\Phi_{a,i}+\frac{\R_{a,i}}{2},\Phi_{a,i}-\frac{\R_{a,i}}{2};0)\,, \label{rhoavars}\ee   and introduce the functional Wigner transform\cite{zubairy,beilok} as a Fourier transform in the relative variable,
\be W[\Phi_{a,i},\Pi_{a,i}] = \int D\R_i \, e^{-i \int d^3 x \Pi_{a,i}(\vx) \,\R_i(\vx)}\,\rho_\phi(\Phi_{a,i}+\frac{\R_{a,i}}{2},\Phi_{a,i}-\frac{\R_{a,i}}{2};0) \,,\label{wigner}\ee which allows us to write (up to a normalization factor)
\be \rho_\phi(\Phi_{a,i}+\frac{\R_{a,i}}{2},\Phi_{a,i}-\frac{\R_{a,i}}{2};0)    = \int D\Pi_{a,i} \, e^{i \int d^3 x  \Pi_{a,i}(\vx) \,\R_{a,i}(\vx)}\, W[\Phi_{a,i},\Pi_{a,i}]
\,, \label{invwig}\ee the variables $\Pi_{a}$ are the momenta conjugate to the variable $\Phi_a$, and $W[\Phi,\Pi]$ yields a probability distribution in ``phase-space'' $\Phi,\Pi$.

Gathering these results together, we now write the generating functional (\ref{Zofj}) in terms of the Keldysh variables (\ref{kelvars}), with the effective action in these variables given by eqn. (\ref{efflanwig}),
 implementing the Wigner transform (\ref{invwig}) and using the representation (\ref{nois}) we obtain
 \bea
  &&  \mathcal{Z}[\mathcal{J}] =
      \int D\Phi_f \int D\R_i\, D\Phi_i\, D\Pi_i \int D\Phi\, D\R\, D\xi\; W[\Phi_i,\Pi_i] \times P[\xi] \times \exp\left\{ i \int dt \sum_{\vec{k}} \Phi_a(\vec{k};t) \J_a(-\vec{k};t) \right\}
    \nonumber \\
    && \times \exp\left\{ -i \int dt \sum_{\vec{k}}   \R_a(-\vec{k};t) \left( \ddot{\Phi}_a(\vec{k};t) + \omega^2_a(k) \Phi_a(\vec{k};t) + \int_0^t \Sigma_{ab}(\vec{k};t-t') \Phi_b(\vec{k},t') dt' - \xi_a(\vec{k};t) \right)   \right\}
    \nonumber \\
    & &  \times \exp\left\{ i \sum_{\vec{k}} \R_{a,i}(-\vec{k}) \left( \Pi_{a,i}(\vec{k}) - \dot{\Phi}_{a,i}(\vec{k}) \right)  \right\}\,,
    \label{Zetafinal}
\eea  where $\omega^2_a(k) = k^2+m^2_a$,  and repeated field indices are summed over. The noise probability distribution function $P[\xi_a]$ is given by
\begin{equation}
    P[\xi_a] = \widetilde{C}\, \prod_{\vec{k}} \exp\left\{ -\frac{1}{2} \int  dt_1 \int  dt_2\, \xi_a(-\vec k;t_1)\,{\mathcal{N} }^{-1}_{ab}(k;t_1-t_2)\,\xi_a(\vec k;t_2) \right\}\,.
    \label{noispdf}
\end{equation}
The generating functional $\mathcal{Z}[\J]$ is the final form of   the time evolved reduced density matrix   after tracing over the bath degrees of freedom. Variational derivatives with respect to the source $\J$ yield  the correlation functions of the Keldysh center of mass variables $\Phi$.

Carrying out the functional integrals over $\R_i(\vec{k})$ and $\R_{\vec{k}}(t)$ yields a more clear form, namely
\begin{align}
    \mathcal{Z}[\mathcal{J}] \propto
   & \int D\Phi_{a,f} \int D\Phi_{a,i}\, D\Pi_{a,i} \int D\Phi_a\, D\xi_a\; W[\Phi_i,\Pi_i] \times P[\xi] \times \exp\left\{ i \int dt \sum_{a,\vec{k}} \Phi_a(\vec{k};t) \J_a(-\vec{k};t) \right\}
    \nonumber \\
    & \times \prod_{\vec{k}} \delta\left[ \ddot{\Phi}_a(\vec{k};t) +  \omega^2_a(k)  \, \Phi_a(\vec{k};t) + \int_0^t \Sigma_{ab}(\vec{k},t-t') \Phi_b(\vec{k};t') dt' - \xi_a(\vec{k};t) \right]\nonumber \\
      & \times \prod_{a,\vec{k}} \delta\left[ \Pi_{a,i}(\vec{k}) - \dot{\Phi}_{a,i}(\vec{k}) \right]\,.
    \label{Zetadelta}
\end{align}

Obtaining expectation values and correlation functions from this generating functional is straightforward:

 \begin{itemize}
 \item The functional delta functions  in (\ref{Zetadelta}) determine   the field configurations that contribute  to the generating functional $\mathcal{Z}[\mathcal{J}]$, these
 are the solutions of the \emph{stochastic} Langevin equation of motion\cite{shuyang} for $\Phi_a(\vec{k};t)$, namely
    \begin{equation}
        \ddot{\Phi}_a(\vec{k};t) + \omega^2_a( {k})\, \Phi_a(\vec{k};t) + \int_0^t \Sigma_{ab}(\vec{k};t-t') \Phi_b(\vec{k};t') dt' = \xi_a(\vec{k};t)\,.\label{langevin}
    \end{equation}
    This equation of motion is retarded as it involves the \emph{retarded} self-energy, thereby defining a causal initial value problem. This is a distinct consequence of the in-in formulation of time evolution.

      \item The expectation value and correlations of the stochastic noise $\xi_a(\vec{k};t)$ are determined by the Gaussian probability distribution $P[\xi_a]$. Introducing the definition $\langle \langle (\cdots) \rangle \rangle$ for averages with $P[\xi_a]$, the Gaussian stochastic noise features the following averages:
    \be
        \langle\langle \xi_a(\vk,t)  \rangle \rangle =0
        ~~;~~
        \langle\langle \xi_a(\vk;t) \xi_b(\vk';t')  \rangle \rangle = \mathcal{N}_{ab}(k;t-t')\,\delta_{\vk,-\vk'} \,. \label{noiscors} \ee Since $P[\xi_a]$ is Gaussian distribution function, higher order correlation functions are obtained by implementing Wick's theorem. This averaging is a manifestation of stochasticity,  establishing a direct relation between non-equilibrium dynamics of quantum open systems and stochastic field theory\cite{calzetta,siggia}.

    \item The stochastic equation of motion (\ref{langevin}) must be solved with the initial conditions
    \begin{equation}
        \Phi_a(\vec{k};t=0) = \Phi_{a,i}(\vec{k}) \qquad;\qquad \dot{\Phi}_a(\vec{k};t=0) = \Pi_{a,i}(\vec{k})\,, \label{inicons}
    \end{equation}
    these initial conditions confirm that $\Pi_{a,i}$ are the canonical momenta conjugate to $\Phi_{a,i}$.       The solution of (\ref{langevin}) is a functional of the variables  $\Phi_{a,i}(\vec{k}),\Pi_{a,i}(\vec{k})$ which are distributed according to the probability distribution function $W[\Phi_{a,i},\Pi_{a,i}]$, which in turn is determined by   the initial density matrix.  This is another manifestation of stochasticity, but now in the distribution of initial conditions.

    We now introduce the notation $\overline{(\cdots)}$ to denote averaging over the initial conditions (\ref{inicons})  with the distribution function $W[\Phi_{a,i},\Pi_{a,i}]$.

\end{itemize}

The solutions of the Langevin equation (\ref{langevin}) $\Phi_a[\vk;t;\xi;\Phi_{a,i};\Pi_{a,i}]$ are \emph{functionals} of the stochastic noise  variables $\xi_a$ and the initial conditions.  Therefore  correlation functions of the original field variables $\phi_a$  in the reduced density matrix correspond to  averaging the products of the solutions over both the initial conditions with the Wigner distribution function $W[\Phi_{a,i},\Pi_{a,i}]$, and the noise with the probability distribution function $P[\xi]$.  We denote such averages  by $\overline{\langle\langle \big( \cdots \big) \rangle\rangle}$ where $\big( \cdots \big)$ is any functional of the initial conditions (\ref{inicons}) and $\xi_a$.

These stochastic averages  yield  the expectation values and correlation functions of functionals of $\Phi_a$ obtained from variational derivatives with respect to $\mathcal{J}_a$.

In appendix (\ref{app:spectralrep}) we provide a non-perturbative spectral Lehmann representation  of the correlation functions $G^{\gtrless}_{ab}(x-y)$ that enter in the definitions of the self-energy (\ref{kernelsigma}) and noise correlation function (\ref{kernelkappa}). The result is that these non-local kernels can be written in a dispersive representation as
\bea \Sigma_{ab}(k;t-t')& = & -i\int \frac{dk_0}{(2\pi)}\,\rho_{ab}(k_0,k)  e^{-ik_0(t-t')}  \label{sigmadis} \\ \mathcal{N}_{ab}(k;t-t') & = & \frac{1}{2}\,\int \frac{dk_0}{(2\pi)}\,\rho_{ab}(k_0,k)\, \coth\big[ \frac{\beta k_0}{2}\big]\,  e^{-ik_0(t-t')} \,, \label{noisedis} \eea where $\rho_{ab}(k_0,k)$ is a $2\times 2$ matrix of spectral densities (see appendix (\ref{app:spectralrep}) for details). The representations (\ref{sigmadis},\ref{noisedis}) are a manifestation of a generalized fluctuation dissipation relation, a consequence of taking the $\chi$ degrees of freedom in thermal equilibrium.

The stochastic equation of motion (\ref{langevin}) with initial conditions (\ref{inicons}) defines an initial value problem whose  solution   is obtained by Laplace transform. Let us define  the Laplace transforms
\bea \widetilde{\Phi}_a(\vk;s) & = &  \int^\infty_0 e^{-st}\,\Phi_a(\vk;t)\,dt \,,\label{laplaA}    \\ \widetilde{\xi}_a(\vk;s) & = &  \int^\infty_0 e^{-st}\,\xi_a(\vk;t)\,dt \,,\label{laplachi} \\ \widetilde{\Sigma}_{ab}(\vk;s) & = &  \int^\infty_0 e^{-st}\,\Sigma_{ab}(\vk;t)\,dt  = -\frac{1}{2\pi} \,\int^{\infty}_{-\infty} \frac{\rho_{ab}(k_0,k)}{k_0-is} dk_0 \,,\label{laplasigma}
\eea where in (\ref{laplasigma}) we used the representation (\ref{sigmadis}). The Laplace transform of the Langevin eqn. (\ref{langevin})  with initial conditions (\ref{inicons}) becomes
\be \mathbb{G}^{-1}_{ab}(k,s)\,\widetilde{\Phi}_b(\vk;s)= \Pi_{a,i}(\vk)+ s\,\Phi_{a,i}(\vk)+ \widetilde{\xi}_a(\vk;s)\,,\label{laplalang}\ee
where
\be \mathbb{G}^{-1}_{ab}(k,s) = (s^2+\omega^2_a(k))\delta_{ab} +\widetilde{\Sigma}_{ab}(\vk;s)\,. \label{inverG} \ee

The solution in real time is obtained by inverse Laplace transform, it is given by
\be \Phi_a(\vk;t) = \Phi^h_a(\vk;t) +  \Phi^{\xi}_a(\vk;t)\,, \label{Asplits} \ee  where $\Phi^h_a(\vk;t); \Phi^{\xi}_a(\vk;t)$ are the homogeneous   and inhomogeneous solutions respectively, namely

\bea \Phi^h_a(\vk;t)  & =  & \dot{\mathcal{G}}_{ab}(k;t)\Phi_{b,i}(\vk)  +  \,\mathcal{G}_{ab}(k;t)\Pi_{b,i}(\vk) \nonumber  \\  \Phi^{\xi}_a(\vk;t) & =  & \int^t_0 \mathcal{G}_{ab}(k;t-t')\,\xi_b(\vk;t')\,dt' \,,\label{realtisol} \eea and repeated indices are summed over. The Green's function is given by
\be \mathcal{G}_{ab}(k;t) = \frac{1}{2\pi i} \int_{\mathcal{C}}   e^{st} \, \mathbb{G}_{ab}(k,s) ds \,, \label{goftsol}\ee   $\mathcal{C}$ denotes the Bromwich contour parallel to the imaginary axis   and to the right of all the singularities   of $\mathbb{G}_{ab}(k,s)$ in the complex s-plane and closing along a large semicircle at infinity with $Re(s)<0$. These singularities correspond to poles and multiparticle branch cuts with $Re(s)<0$, thus the contour runs parallel to the imaginary axis $s= i(\nu -i \epsilon)$, with $-\infty \leq \nu \leq \infty$ and $\epsilon \rightarrow 0^+$. Finally, changing variables $\nu = -\omega$,   we obtain
\be \mathcal{G}_{ab}(k;t) =   \int^{\infty}_{-\infty}  \mathbb{G}_{ab}(k,s=-i\omega+\epsilon)\, {e^{-i\omega t}} \,\frac{d\omega}{2\pi}\,, \label{Goftfin}\ee and for $t>0$ the integration contour is closed in the lower half $\omega$-plane. We obtain the Green's function $\mathbb{G}_{ab}(k,s)$ by following the steps  in appendix (\ref{app:green}).  Without loss of generality we consider $m^2_1 \geq m^2_2$ and define
\bea \overline{\M}(k,s) & = & s^2  + \frac{1}{2}\Big[ \omega^2_1(k)+\omega^2_2(k)+\widetilde{\Sigma}_{11}(k,s)+\widetilde{\Sigma}_{22}(k,s) \Big] \,\label{invgM}\\
\mathcal{D}(k,s) & = & \Bigg[\Big(\omega^2_1(k)-\omega^2_2(k)+\widetilde{\Sigma}_{11}(k,s)-\widetilde{\Sigma}_{22}(k,s)\Big)^2+ 4 \widetilde{\Sigma}_{12}(k,s)\widetilde{\Sigma}_{21}(k,s) \Bigg]^{1/2}\,\label{invgD}\\
\oalpha(k,s) & = & \frac{1}{\mathcal{D}(\vk,s)}\Big(\omega^2_1(k)-\omega^2_2(k)+\widetilde{\Sigma}_{11}(\vk,s)-\widetilde{\Sigma}_{22}(\vk,s)\Big)\,\label{alfainvg}\\
\obeta(k,s) & = & \frac{2\,\widetilde{\Sigma}_{12}(k,s)}{\mathcal{D}(\vk,s)}~;~ \ogamma(k,s) =\frac{2\,\widetilde{\Sigma}_{21}(k,s)}{\mathcal{D}(k,s)} \,,\label{betgaminvg}
\eea with the property
\be \oalpha^2(k,s)+\obeta(k,s)\ogamma(k,s) = 1 \Rightarrow \alpha(k,s)= \sqrt{1-\obeta(k,s)\,\ogamma(k,s)}\,,   \label{nuid}\ee  where we used the same argument leading to eqn.  (\ref{alfi}) for the choice of sign for $\oalpha$.

In terms of these variables , $\mathbb{G}^{-1}_{ab}(k,s)$ has the same form as eqn. (\ref{mtxM2}) in appendix (\ref{app:green}), yielding
\be  \mathbb{G}(k,s) =  \frac{\mathbb{P}_-(k,s)}{\overline{\M}(k,s)  - \frac{ \mathcal{D}(k,s)}{2} } +  \frac{\mathbb{P}_+(k,s)}{\overline{\M}(k,s)  + \frac{ \mathcal{D}(k,s)}{2} }~~;~~ \mathbb{P}_\pm = \frac{1}{2} \Big(\mathbf{1}\pm \mathbb{R} \Big) \,,\label{greenfin}   \ee with
 \be \mathbb{R} = \left(
                                                                           \begin{array}{cc}
                                                                             \oalpha & \obeta \\
                                                                             \ogamma & -\oalpha \\
                                                                           \end{array}
                                                                         \right)   ~~;~~ \mathbb{R}^2 = \mathbf{1} \Rightarrow \mathbb{P}^2_\pm(k,s)=\mathbb{P}_\pm(k,s)\,. \label{Geffin}\ee The analytic continuation of the self energies is
\be \widetilde{\Sigma}_{ab}(k,s=-i\omega+\epsilon)= \frac{1}{2\pi}\Bigg[\int^\infty_{-\infty} \mathcal{P}\,\, \frac{\rho_{ab}(k_0,k)}{\omega-k_0  }\,dk_0 -i\pi \,\rho_{ab}(\omega,k) \Bigg]\,.\label{sigconti} \ee

Upon analytic continuation to $\omega$ the Green's function (\ref{Goftfin}) becomes
\be \mathcal{G}_{ab}(k;t) =   -\int^{\infty}_{-\infty}  \Bigg[  \frac{\mathbb{P}_-(k,\omega)}{(\omega+i\epsilon)^2  - \Omega^2_-(k,\omega) } +  \frac{\mathbb{P}_+(k,\omega)}{(\omega+i\epsilon)^2  - \Omega^2_+(k,\omega)} \Bigg]_{ab}  \, {e^{-i\omega t}} \,\frac{d\omega}{2\pi}\,,\label{gomegt}  \ee
with
\be \Omega^2_\pm(k,\omega) = \frac{1}{2}\,\Big[ \omega^2_1(k)+\omega^2_2(k)+\widetilde{\Sigma}_{11}(k,\omega)+\widetilde{\Sigma}_{22}(k,\omega) \pm \D(k,\omega) \Big] \,,\label{bigome} \ee where the functions of $\omega$ are understood as the functions of $s$ upon   analytic continuation $s=-i\omega+\epsilon$, keeping the same name for the functions to simplify notation.

The form of the Green's function is similar to eqn. (\ref{iLTw}) of the previous section, with important differences: whereas the denominators in eqn. (\ref{iLTw}) are linear in $\omega$, therefore each term features only one pole, the denominators in (\ref{gomegt}) are quadratic in $\omega$ implying that each term features two poles. This discrepancy has a simple explanation: the set of amplitude equations leading up to (\ref{iLTw}) describe the evolution of \emph{single particle} states, whereas the effective field theory yields the time evolution of the density matrix in the field basis, a real scalar field describes positive frequency particle states and negative frequency antiparticle states. Even in absence of perturbations, the propagator has two poles yielding the time evolution $e^{\mp i \omega(k) t}$ for the amplitudes. Furthermore, the self energies $\widetilde{\Sigma}_{ab}(k,\omega)$ have dimensions of energy squared, unlike the quantities $\Delta_{ab}$ in the previous section that feature dimensions of energy.

The complex poles in the Green's functions are at
\be \omega^2_\pm = \Omega^2_{\pm}(k,\omega_\pm)\,, \label{polis}\ee namely
\bea \omega^{(\pm)}_+ & = & (\pm) \, \Omega_+(k,\omega^{(\pm)}_+) \,,\label{omepluspm}\\
\omega^{(\pm)}_- & = & (\pm)\, \Omega_-(k,\omega^{(\pm)}_-) \,,\label{omeminpm}
\eea where the superscripts $(\pm)$ denote the two roots of (\ref{omepluspm},\ref{omeminpm}) for each subscript label $+,-$ corresponding to the signs in of $\Omega^2_\pm$ in (\ref{bigome}). These roots define the complex frequencies of the  \emph{quasinormal} modes.

Consistently with perturbation theory,  we assume    that $\omega^2_{1,2}(k) \gg \widetilde{\Sigma}_{ab}(k,\omega)\propto H^2_I$ allowing to implement  Breit-Wigner and narrow width approximations to the propagators. Just as in the case discussed in the previous section, the validity of the Breit-Wigner approximation relies on weak coupling, in particular that the distance between the real
part of the poles and thresholds is much larger than the half-width of the resonance. This criterion holds for both the non-degenerate and nearly degenerate cases, because in
the latter the condition of near degeneracy is that $\omega^2_a \gg \tilde{\Sigma}_{ab}$ and $\omega^2_1- \omega^2_2 \lesssim \tilde{\Sigma}_{ab}$. This approximation describes exponential relaxation valid in the intermediate time scale as discussed above.

In these approximations,   we expand around each pole in the denominators in (\ref{gomegt}), namely
\be \omega = \omega^{(+)}_+ + (\omega-\omega^{(+)}_+)~~;~~  \,\Omega_+(k,\omega) =  \,\Omega_+(k,\omega^{(+)}_+)+ (\omega-\omega^{(+)}_+)\,\frac{1}{2\,\omega^{(+)}_+}\,\frac{d}{d\omega}\Omega^2_+(k,\omega)\big|_{\omega=\omega^{(+)}_+}+\cdots \label{expapo}\ee and similarly for each of the other poles. Using the pole condition (\ref{polis}) yields the general form of the Green's function,
\be  \mathcal{G}_{ab}(k;t) = \mathcal{G}_{ab,+}(k;t)+ \mathcal{G}_{ab,-}(k;t)\,,\label{Gtotnm}\ee where each term corresponds to the contribution of the individual quasinormal modes corresponding to the subscripts $\pm$, namely

\bea \mathcal{G}_{ab,+}(k;t) & = &  i\Bigg[\mathcal{Z}^{(+)}_{+}\,\frac{e^{-i\omega^{(+)}_+ t}}{2\omega^{(+)}_+}\,\mathbb{P}_+( \omega^{(+)}_+)+\mathcal{Z}^{(-)}_{+}\,\frac{e^{-i\omega^{(-)}_+ t}}{2\omega^{(-)}_+}\,\mathbb{P}_+( \omega^{(-)}_+) \Bigg]_{ab} \label{figriplu} \\ \mathcal{G}_{ab,-}(k;t)& = &  i\Bigg[\mathcal{Z}^{(+)}_{-}\,\frac{e^{-i\omega^{(+)}_- t}}{2\omega^{(+)}_-}\,\mathbb{P}_-( \omega^{(+)}_-)+\mathcal{Z}^{(-)}_{-}\,\frac{e^{-i\omega^{(-)}_- t}}{2\omega^{(-)}_-}\,\mathbb{P}_-( \omega^{(-)}_-) \Bigg]_{ab}\,.\label{figrimin} \eea The wavefunction renormalization constants   are given by
\be \mathcal{Z}^{(\pm)}_{+} = \Bigg[1- \frac{d}{d\omega}\frac{\Omega^2_+(k,\omega)}{2\omega^{(\pm)}_+}\Big|_{\omega=\omega^{(\pm)}_{+}} \Bigg]^{-1}~~;~~
\mathcal{Z}^{(\pm)}_{-} = \Bigg[1- \frac{d}{d\omega}\frac{\Omega^2_-(k,\omega)}{2\omega^{(\pm)}_-}\Big|_{\omega=\omega^{(\pm)}_{-}} \Bigg]^{-1}\,. \label{zeta7}\ee The Green's function for each quasinormal mode features both positive ($\omega^{(+)}$) and negative ($\omega^{(-)}$) frequency contributions.

  This is the general result for the Green's function, again displaying the four poles: positive and negative frequency for each quasinormal mode,  with the associated wave function renormalization constants arising from the residues at the poles in the Breit-Wigner approximation.

  Note that while the result  (\ref{solLT}) only features positive frequency components,  the Green's function (\ref{Gtotnm}) features both positive ($\omega^{(+)} $)  and negative ($\omega^{(-)} $) frequency components.    As discussed above the origin of this difference is that whereas the Weisskopf-Wigner formulation, upon which the (LOY) theory is based,  describes the time evolution of single particle (positive frequency) or antiparticles (negative frequency) amplitudes, the Green's function in the effective field theory describes the propagation of \emph{fields} that include both positive and negative frequencies and describe the \emph{quasi} normal modes of propagation as a consequence of mixing and decay. Although this expression looks cumbersome in its index structure, we clarify again: the superscripts $(\pm)$ refer to the positive (particle) and negative (antiparticle) frequencies, the subscripts $\pm$ refer to the two (quasi) normal modes from mixing, corresponding to the $\pm$ in eqn. (\ref{bigome}).

With the purpose of comparison with the (LOY) theory, we focus on  the same possible scenarios as in the previous section.

\vspace{1mm}

 \textbf{  Non-degenerate:   $m^2_1- m^2_2 \gg \widetilde{\Sigma}_{ab}$.} In this case it follows that
 \be \D(k,\omega) \simeq  \omega^2_1(k)-\omega^2_2(k)+\widetilde{\Sigma}_{11}(k,\omega)-\widetilde{\Sigma}_{22}(k,\omega)\,,\label{Dnondeg} \ee
  yielding up to second order in the couplings
 \bea \Omega^2_+(k,\omega) & = & \omega^2_1(k) + \wS_{11}(k,\omega) \,,\label{Omepnd}\\
\Omega^2_-(k,\omega) & = & \omega^2_2(k) + \wS_{22}(k,\omega) \,. \label{Omemnd} \\
\oalpha(k,\omega) & \simeq & 1 +\mathcal{O}(\wS^2) \,\label{oalf1}\\
\obeta(k,\omega) & \simeq & \frac{2\,\wS_{12}(k,\omega)}{m^2_1-m^2_2}\ll 1\,\label{obet1}\\
\ogamma(k,\omega) & \simeq & \frac{2\,\wS_{21}(k,\omega)}{m^2_1-m^2_2}\ll 1 \,.\label{ogam1}\eea Therefore, up to second order in couplings the complex  poles are at

\bea \omega^{(\pm)}_+(k) & = &  \pm\omega_{+r}(k) -i\,\frac{\Gamma^{(\pm)} _+(k)}{2} \,,\label{polespm1nd} \\
\omega^{(\pm)}_-(k) & = &  \pm \omega_{-r}(k) -i\,\frac{\Gamma^{(\pm)}_-(k)}{2}  \,,\label{polespm2nd}\eea where
\be \omega_{+r}(k)= \omega_{1}(k)+\delta_+(k) ~~;~~ \omega_{-r}(k)= \omega_{2}(k)+\delta_-(k)\,, \ee are the renormalized frequencies of each quasinormal mode, and to leading order

\bea \delta_{+}(k) & = &  \frac{\mathrm{Re}\wS_{11}(k,\omega_1(k))}{2\,\omega_1(k)}~~;~~ \delta_{-}(k)   =   \frac{\mathrm{Re}\wS_{22}(k,\omega_2(k))}{2\,\omega_2(k)}\,,\label{deltis} \\
  \Gamma^{(\pm)}_+(k)  & = & (\pm)\frac{\rho_{11}(\pm\omega_1(k),k)}{  2\omega_1(k)}~~;~~ \Gamma^{(\pm)}_-(k)    =  (\pm) \frac{\rho_{22}(\pm\omega_2(k),k)}{2\omega_2(k)}\,. \label{gamis} \eea

 To obtain the above results we used the property $\rho_{aa}(-\omega,k) = -\rho_{aa}(\omega,k)$  (no sum over $a$) for the diagonal matrix elements of the spectral density      (see eqn. (\ref{oddro}) in appendix (\ref{app:spectralrep})). The contributions $\delta_{\pm}(k)$ are renormalization of the bare frequencies $\omega_{1,2}$ respectively.

  The Green's function (\ref{Gtotnm}) with (\ref{figriplu},\ref{figrimin}) is given to leading order  in this case by

\bea \mathcal{G}_{ab,+}(k;t) & = &  \frac{i }{2\,\omega_{+r}}\Bigg[\mathcal{Z}^{(+)}_{+}\, e^{-i\omega_{+r}\, t}\,e^{-\frac{\Gamma^+_+}{2}\,t} \,\mathbb{P}_+( \omega_1)-\mathcal{Z}^{(-)}_{+}\,e^{i\omega_{+r} t}\,e^{-\frac{\Gamma^-_+}{2}\,t}\,\mathbb{P}_+(- \omega_1) \Bigg]_{ab} \label{figriplund} \\ \mathcal{G}_{ab,-}(k;t) & = &  \frac{i }{2\,\omega_{-r}}\Bigg[\mathcal{Z}^{(+)}_{-}\, e^{-i\omega_{-r}\, t}\,e^{-\frac{\Gamma^+_-}{2}\,t} \,\mathbb{P}_-( \omega_2)-\mathcal{Z}^{(-)}_{-}\,e^{i\omega_{-r} t}\,e^{-\frac{\Gamma^-_-}{2}\,t}\,\mathbb{P}_-(- \omega_2) \Bigg]_{ab} \,,\label{figriminnd} \eea with the projection operators   given to leading order in the couplings by
\bea
\mathbb{P}_+( \pm \omega_1) & = & \left(
                                    \begin{array}{cc}
                                      1 & \frac{1}{2}\,\obeta(k,\pm\omega_1) \\
                                      \frac{1}{2}\,\ogamma(k,\pm\omega_1) & 0 \\
                                    \end{array}
                                  \right) \label{Ppnd}\\
 \mathbb{P}_-( \pm \omega_2) & = & \left(
                                    \begin{array}{cc}
                                      0 & -\frac{1}{2}\,\obeta(k,\pm\omega_2) \\
                                      -\frac{1}{2}\,\ogamma(k,\pm\omega_2) & 1 \\
                                    \end{array}
                                  \right) \,. \label{Pmnd}
\eea
Since in this non-degenerate case $\obeta,\ogamma \propto \Sigma$ (see eqns. (\ref{obet1},\ref{ogam1})) the off diagonal terms are perturbatively small, for these the wave function renormalization constants can be set $\mathcal{Z} \simeq 1$ to leading order. This result agrees with those obtained in ref.\cite{mesonmix} for the case of pion-axion mixing.

\vspace{1mm}

\textbf{(Nearly)-degenerate: $m^2_1-m^2_2 \lesssim \widetilde{\Sigma}_{ab}$.} In this case, it is convenient to define
\be \Omega^2_{\pm}(k,\omega) = \overline{\omega}^{\,2}(k)+ \mathcal{E}_{\pm}(k,\omega)\,,\label{bigomedeg}\ee with
\be \overline{\omega} (k) = \frac{1}{\sqrt{2}}\,\big[\omega^2_1(k)+\omega^2_2(k)\big]^{1/2} \gg \wS_{ab},\D \,,\label{omeav}\ee and
\be   \mathcal{E}_{\pm}(k,\omega) = \frac{1}{2}\Big[\wS_{11}(k,\omega)+\wS_{22}(k,\omega)\pm \mathcal{D}(k,\omega)\Big] \ll \overline{\omega}^{\,2}\,. \label{bigE}\ee with  $\wS_{ab}$ and $\mathcal{D}$ of the same order. In this  case the complex poles are at
\bea \omega^{(\pm)}_+(k) & = &  \pm {\omega}_{+r}(k) -i\,\frac{\Gamma^{(\pm)}_+(k)}{2} \,,\label{polespm1} \\
\omega^{(\pm)}_-(k) & = &  \pm {\omega}_{-r}(k) -i\,\frac{\Gamma^{(\pm)}_-(k)}{2} \,,\label{polespm2}\eea where
\be  {\omega}_{\pm r}(k) = \overline{\omega}(k)+\delta _{\pm}(k)\,,\ee are the renormalized (nearly degenerate) frequencies, with
\be  \delta_\pm(k)   =    \frac{\mathrm{Re}\Big[\mathcal{E}_\pm(k,\pm\overline{\omega}(k))\Big]}{2\,\overline{\omega}(k)}  ~~;~~
  \Gamma^{(\pm)}_\pm(k)   =    (\mp)\frac{\mathrm{Im}\Big[\mathcal{E}_\pm(k,\pm\overline{\omega}(k))\Big]}{  \overline{\omega}(k)}\, ,\label{delgamsdeg}\ee  and both are of quadratic order in the couplings.

  We assume that the decay rates $\Gamma^{(\pm)}_\pm$ are all positive for stability, further properties of   $\delta_\pm ~; ~ \Gamma^{(\pm)}_\pm$ will depend on the specific details of the self-energies $\wS_{ab}$ which in turn depend   on the type of operators $\mathcal{O}_{a}$.  In this case  $\omega_{+r}-\omega_{-r} \simeq \wS$, namely the difference in the quasinormal mode frequencies are of quadratic order in the couplings.    To leading order in the couplings the Green's functions (\ref{figriplu},\ref{figrimin}) in this nearly degenerate case are given by

 \bea \mathcal{G}_{+}(k;t) & = &  \frac{i}{2\, {\omega}_{+r}}\Bigg[\mathcal{Z}^{(+)}_{+}\, e^{-i\,  {\omega}_{+r}\, t}\,e^{-i \frac{\Gamma^+_+}{2}t } \,\mathbb{P}_+( {\omega}_{+r})-\mathcal{Z}^{(-)}_{+}\,e^{i {\omega}_{+r}\,t}\,e^{-i \frac{\Gamma^-_+}{2}t } \,\mathbb{P}_+(-  {\omega}_{+r}) \Bigg] \label{figripludeg} \\ \mathcal{G}_{-}(k;t)& = & \frac{i}{2\, {\omega}_{-r}}\Bigg[\mathcal{Z}^{(+)}_{-}\, e^{-i\,  {\omega}_{-r}\, t}\,e^{-i \frac{\Gamma^+_-}{2}t } \,\mathbb{P}_-( {\omega}_{-r})-\mathcal{Z}^{(-)}_{-}\,e^{i {\omega}_{-r}\,t}\,e^{-i \frac{\Gamma^-_-}{2}t } \,\mathbb{P}_-(-  {\omega}_{-r}) \Bigg]  \,,\label{figrimindeg} \eea where  all matrix elements of $\mathbb{P}_{\pm}$ are of $\mathcal{O}(1)$.

\vspace{1mm}

\subsection{Expectation values and  correlation functions:}\label{subsecexp}

We seek to obtain expectation values and correlation functions of $\phi_a$ in the reduced density matrix. In particular we focus on equal time correlation functions, if asymptotically at long time these become time independent, this is a signal of the emergence of a stationary state, from which we can assess if the fields reach thermal equilibration with the bath. Furthermore, off-diagonal equal time correlation functions will also inform on the emergence and long-time survivability of \emph{coherence}.
Therefore we must relate these to the averages of the center of mass Keldysh fields $\Phi_a$. To establish this relation, we begin with the path integral representations for the forward and backward time evolution operators (\ref{evolrhot}, \ref{piforward}, \ref{piback}) which show that $\phi^+_a$ are associated with $U(t)$ and $\phi^-_a$ with $U^{-1}(t)$, hence it follows that    the expectation value of the fields in the full density matrix  is given by
\be  \langle \phi_a(\vx,t) \rangle   =   \mathrm{Tr}\phi^+_a(\vx,t)\,\hat{\rho}(0)=   \mathrm{Tr}  \hat{\rho}(0)\,\phi^-_a(\vx,t)=  \mathrm{Tr} \Phi_a(\vx,t)\, {\hat{\rho}(0)}(0) = \overline{\langle \langle \Phi_a(\vx,t) \rangle \rangle} \,,\label{averageA} \ee
whereas
\be \mathrm{Tr} \R_a(\vx,t)\,\hat{\rho}(0) =0 \label{aveR}\,. \ee

Similarly, correlation functions  in the forward, backward and mixed forward-backward branches, are given by

 \bea && \mathrm{Tr}\phi^+_a(\vk;t)\phi^+_b(\vk';t')\,\hat{\rho}(0) \equiv \mathrm{Tr}\,T\big(\phi_a (\vk;t)\phi_b (\vk';t'))\, \hat{\rho}(0) \nonumber \\
&& \mathrm{Tr} \phi^-_a(\vk;t)\phi^-_b(\vk';t')\,\hat{\rho}(0) \equiv \mathrm{Tr} \hat{\rho}(0)\,\widetilde{T}\big( \phi_a(\vk;t)\phi_b(\vk';t')\big) \nonumber \\
 && \mathrm{Tr} \phi^+_a(\vk;t)\phi^-_b(\vk';t')\,\hat{\rho}(0) \equiv  \mathrm{Tr} \phi_a(\vk;t)\, \hat{\rho}(0)\, \phi_b(\vk';t') = \mathrm{Tr}\,\phi_b(\vk';t') \, \phi_a(\vk;t)\, \hat{\rho}(0)  \,,\label{dict}\eea where $T,\widetilde{T}$ are the time ordering and anti-time ordering symbols respectively.  Using the relations (\ref{dict}) it is straightforward to confirm that
\be \mathrm{Tr}\,\Phi_a(\vx,t)\Phi_b(\vx',t')\,\hat{\rho}(0) \equiv \frac{1}{2}\,\mathrm{Tr}\Big( \phi_a(\vx,t)\phi_b(\vx',t')+\phi_b(\vx',t')\phi_a(\vx,t)\Big)\,\hat{\rho}(0)\,.\label{ide} \ee

Upon obtaining the functional solutions of eqn. (\ref{langevin}) our objective is to obtain the connected equal time correlation functions
\be \langle \phi_a(t) \phi_b(t) \rangle_c= \mathrm{Tr} \hat{\rho}(0) \phi_a(t) \phi_b(t) -  \mathrm{Tr} \hat{\rho}(0) \phi_a(t) \, \mathrm{Tr}\hat{ \rho}(0)   \phi_b(t)\,.  \label{conncorr} \ee
  and the population for each field component of wavevector $k$, namely
\be n_a(k;t) = \frac{1}{2\omega_a(k)}\,\mathrm{Tr}\hat{\rho}(0)\, \Big[\dot{\phi}_a(\vk;t)\dot{\phi}_a(-\vk;t)+\omega^2_a(k) \, {\phi}_a(\vk;t) {\phi}_a(-\vk;t)\Big]  - \frac {1}{2} ~~~~(\mathrm{no~sum~over~a}) \,.\label{numbs}\ee

Establishing contact with the dynamics of the density matrix of two level systems\cite{zubairy},   the off-diagonal components of the connected correlation function (\ref{conncorr})  are a manifestation of  \emph{coherence}. If initially the fields are uncorrelated, the off diagonal components of the correlation function vanish. Therefore if upon time evolution these are non-vanishing, these off-diagonal correlations between the two fields are a consequence of coherence induced by the indirect mixing through the interactions of the field with the bath.

With the definition of the Keldysh center of mass field variables $\Phi_a$ (\ref{kelvars}) and the relations (\ref{dict},\ref{averageA},\ref{aveR}) we find that the equal time connected correlation function (\ref{conncorr}) is given by
\be \langle \phi_a(t) \phi_b(t) \rangle_c= \overline{\langle\langle \Phi_a(t) \Phi_b(t) \rangle \rangle}- \overline{\langle\langle \Phi_a(t)   \rangle \rangle} ~~ \overline{\langle\langle   \Phi_b(t) \rangle \rangle}\,.\label{conncorrfin2}\ee To obtain the population for each field (\ref{numbs})we now introduce
\be \mathbb{C}^>_a(k;t,t') = \mathrm{Tr} \phi^{-}_a(\vk;t) \phi^{+}_a(-\vk;t')  {\rho}(0)~~;~~ \mathbb{C}^<_a(k;t,t') = \mathrm{Tr} \phi^{-}_a(\vk;t') \phi^{+}_a(-\vk;t) {\rho}(0)\,, \label{gis}\ee  the populations   the wavevector $\vk$ component of each field $\phi_a$ (\ref{numbs}) become
\be n_a(k;t) = \frac{1}{4\,\omega_a(k)}\,  \Bigg(\frac{\partial}{\partial t}\frac{\partial}{\partial t'} +\omega^2_a(k)  \Bigg)\Bigg[\mathbb{C}^>_a(k;t,t')+\mathbb{C}^<_a(k;t,t') \Bigg]_{t=t'}-\frac{1}{2} \,. \label{number} \ee   Using the definition (\ref{kelvars}) and the relations (\ref{dict}) it is straightforward to show that this symmetrized product yields
\bea n_a(k;t)  & = &  \frac{1}{2\omega_a(k) } \mathrm{Tr}{\rho}(0) \Bigg(\dot{\Phi}_a(\vk;t)\dot{\Phi}_a(-\vk;t)+ \omega^2_a(k) {\Phi}_a(\vk;t) {\Phi}_a(-\vk;t)  \Bigg)-\frac{1}{2}\,   \nonumber \\ & = & \frac{1}{2\omega_a(k)}\,\Bigg\{\overline{\langle \langle \dot{\Phi}_a(\vk;t)\dot{\Phi}_a(-\vk;t)\rangle \rangle}+ \omega^2_a(k) \,\overline{\langle \langle {\Phi}_a(\vk;t) {\Phi}_a(-\vk;t)  \rangle \rangle}\Bigg\}-\frac{1}{2} \,.\label{eneA} \eea
The corollary of this analysis is that we can obtain the connected correlation functions   and the populations of the fields $\phi_{1,2}$ by obtaining the solutions of the Langevin equation of motion (\ref{langevin}) with initial conditions (\ref{inicons}) and taking the averages over the initial conditions and noise described above.

Armed with the solution of the Langevin equations (\ref{Asplits},\ref{realtisol})), the above results, and the general form for the Green's function (\ref{Gtotnm}), in terms of the Green's functions for the quasinormal modes (\ref{figriplu},\ref{figrimin}),  we can now study the expectation values, connected correlation functions (\ref{conncorrfin2}) and populations (\ref{eneA}). The solution (\ref{Asplits},\ref{realtisol})) along with the averages (\ref{noiscors}) yield for the spatial Fourier transform of the fields
\be \langle \phi_a(\vk,t) \rangle = \dot{\mathcal{G}}_{ab}(k;t)\,\overline{\Phi _{b,i}(\vk)}  +  \,\mathcal{G}_{ab}(k;t)\,\overline{\Pi _{b,i}(\vk)} \,.\label{avefie}\ee

Similarly,    the connected correlation functions (\ref{conncorrfin2}) are
\be \langle \phi_a(\vk,t) \phi_b(-\vk,t) \rangle_c= \overline{  \Phi^h_a(\vk,t) \Phi^h_b(-\vk,t)}- \overline{  \Phi^h_a(\vk,t) } ~~ \overline{    \Phi^h_b(-\vk,t)  }+ \langle \langle \Phi^\xi_a(\vk,t) \Phi^\xi_b(-\vk,t)\rangle \rangle \,,\label{conec}\ee where $\Phi^h,\Phi^\xi$ are given by (\ref{realtisol}). These are general results for expectation values and correlation functions, from which we can obtain their time evolution.

\subsection{Time evolution, thermalization and bath-induced coherence.}\label{subsec:timeevol}

Taken together, the results (\ref{avefie},\ref{conec})   inform important aspects for the time evolution of expectation values and correlation functions:

\textbf{I:)} Even when initially only one of the fields, for example $\phi_1$, features an expectation value, the off-diagonal components of the Green's functions determined by the projector operators $\mathbb{P}_\pm$ in (\ref{figriplu},\ref{figrimin}) induce a non-vanishing expectation value for the other field, in this case $\phi_2$. This phenomenon has been noticed in ref.\cite{mesonmix} in the case of axion-pion mixing. In the (LOY) theory discussed in the previous section, a similar feature emerges at the level of the\emph{ amplitudes} of the single particle states $\ket{\phi_1}, \ket{\phi_2}$:   for example if the initial amplitudes are $C_1(0) \neq 0; C_2(0) =0$, upon time evolution a non-vanishing amplitude $C_2(t)$ is induced as a consequence of  mixing. The off diagonal components of $\mathcal{G}_{ab}$ are a   consequence of the off-diagonal components of the self-energy matrix and a direct manifestation of the couplings of the fields to correlated operators of the bath degrees of freedom, namely ``indirect'' mixing.

\textbf{II:)} A similar phenomenon emerges for the connected correlation function (\ref{conec}). Even if   the fields $\phi_{1,2}$ are initially uncorrelated, a non-vanishing correlation emerges from   the off-diagonal components of the noise correlation function that determines the last term in (\ref{conec}). We refer to the emergence of non-vanishing correlations as bath-induced coherence, referring as coherence to the off diagonal connected correlation functions of the field in agreement with the description of the time evolution of a density matrix  in two level systems\cite{zubairy}.

\textbf{III:)} The off-diagonal components of the projectors $\mathbb{P}_\pm$ are perturbatively small $\mathcal{O}(\Sigma)$ in the non-degenerate case, whereas they are of $\mathcal{O}(1)$ in the nearly degenerate case. In turn this implies that the induced expectation values and coherence are perturbatively small in the non-degenerate case, in agreement with the results in ref.\cite{mesonmix}, but  are of $\mathcal{O}(1)$ in the nearly degenerate case. This  expectation is confirmed by the analysis below.

 The first two terms in (\ref{conec}) decay exponentially because the Green's functions do, and depend explicitly on the initial conditions. The last term is independent of the initial conditions, it is completely determined by the noise term  induced by the bath degrees of freedom and, as we show below,  survives in the long time limit, hence determining the approach to a stationary state.

 We now focus on this last term, which upon using the noise correlation function (\ref{noiscors},\ref{noisedis}) is given by
\be  \langle \langle \Phi^\xi_a(t) \Phi^\xi_b(t)\rangle \rangle = \frac{1}{2} \int^\infty_{-\infty} \frac{dk_0}{2\pi}\,\Bigg[\int^t_0 \mathcal{G}_{ac}(k,\tau)\,e^{i k_0 \,\tau}\,  d\tau\Bigg] \Bigg[\int^t_0 \mathcal{G}_{bd}(k,\tau)\,e^{-i k_0 \,\tau}\, d\tau\Bigg] \, \rho_{cd}(k_0)\,\coth\Big[\frac{\beta k_0}{2} \Big]\,.\label{noiscor7}\ee Each of the $\mathcal{G}'s$ in this expression is a sum of the Green's functions of each quasinormal mode given by eqns. (\ref{Gtotnm},\ref{figriplu},\ref{figrimin}), therefore each $\mathcal{G}$ features four terms,  hence there are altogether sixteen terms in (\ref{noiscor7}). Because $\rho_{cd}$ is of second order in couplings, we will focus on   the terms that are of $\mathcal{O}(1)$ in these couplings, these arise from the terms that feature small denominators of second order in the couplings that compensate the numerator $\rho_{cd}$.

Each of the terms in $\mathcal{G}$ features exponentials of the form $e^{-i( \pm W -i\Gamma/2)t}$ where $W$ stands for the real part of the quasinormal mode
frequencies, the $\pm$ describing the positive and negative frequency components, and $\Gamma>0$ stands generically for the decay rate of these modes, therefore the
time integral of such typical term in the first bracket in (\ref{noiscor7}) yields
\be \int^t_0 e^{i(k_0 \mp W +i\Gamma/2)\tau} d\tau = \frac{e^{i(k_0 \mp  W +i\Gamma/2)t}-1}{i(k_0 \mp  W +i\Gamma/2)} \,,\label{integno}\ee and for the second bracket there is a similar generic contribution but with $k_0\rightarrow -k_0$. Obviously as $t\rightarrow \infty$ these contributions remain non-vanishing, confirming that the noise contribution to the correlation functions and coherences (off diagonal) remain finite in the long time limit. The $k_0$ integral in (\ref{noiscor7}) is dominated by the   poles in the complex $k_0$ plane. To identify these, consider the product of the positive frequency contribution for the first $\mathcal{G}$ with the negative frequency contribution of the second $\mathcal{G}$ for the \emph{same quasinormal mode}, for example that of frequency $\omega_{+r}$, such a term is proportional to the product
\be \Bigg[\frac{e^{i(k_0 -  \omega_{+r} +i\frac{\Gamma^+_+}{2})\,t}-1}{i(k_0 -  \omega_{+r} +i\frac{\Gamma^+_+}{2})}\Bigg]\,\Bigg[\frac{e^{-i(k_0 -  \omega_{+r} -i\frac{\Gamma^-_+}{2})\,t}-1}{(-i)(k_0 -  \omega_{+r} -i\frac{\Gamma^-_+}{2})}\Bigg]= \frac{1+e^{-(\Gamma^+_++\Gamma^-_+) t/2}-e^{i(k_0 -  \omega_{+r} +i\frac{\Gamma^+_+}{2})\,\,t}-e^{-i(k_0 -  \omega_{+r} -i\frac{\Gamma^-_+}{2})\,t}}{(k_0 -  \omega_{+r} +i\frac{\Gamma^+_+}{2})(k_0 -  \omega_{+r}-i\frac{\Gamma^-_+}{2})} \,,\label{direc}  \ee we refer to these as \emph{direct} terms, there are \emph{two} for each quasinormal mode. These terms feature poles at $k_0 = \omega_{+r} \pm i\frac{\Gamma^+_+}{2}$ yielding for the integral a contribution proportional to
\be \propto 4\times 2\pi\,\mathbb{P}_+(\omega_{+r})_{ac}\frac{\rho_{cd}(\omega_{+r})}{\Gamma^+_++\Gamma^-_+}\,\mathbb{P}_+(-\omega_{+r})_{bd}\,(1+2\,n(\omega_{+r}))\,(1-e^{-(\Gamma^+_++\Gamma^-_+) \,t/2})\,, \label{direint}\ee where $n(\omega)$ is the Bose-Einstein distribution function with energy $\omega$. In this expression we kept the leading order terms, neglecting wave function renormalization constants, and with $\rho_{cd}(\omega)$ and the $\coth(\omega)$ evaluated at $\omega_{+r}$, namely the real part of the frequency of the quasinormal mode, since $\Gamma^{\pm}_+$ are of second order in couplings. An important aspect of this contribution is that it is of $\mathcal{O}(1)$, because $\rho$ and $\Gamma$ are both of second order in the couplings.
Now consider the positive (or negative)  frequency contributions of the \emph{same} quasinormal modes in both brackets, for example, for the positive frequency  whose time integrals yield a term proportional to
\bea && \Bigg[\frac{e^{i(k_0 -  \omega_{+r} +i\frac{\Gamma^+_+}{2})\,t}-1}{i(k_0 -  \omega_{+r} +i\frac{\Gamma^+_+}{2})}\Bigg]\,\Bigg[\frac{e^{-i(k_0 +  \omega_{+r} -i\frac{\Gamma^+_+}{2})\,t}-1}{(-i)(k_0 +  \omega_{+r} -i\frac{\Gamma^+_+}{2})}\Bigg] \nonumber \\  & = &\frac{1+e^{-2i\omega_{+r}\,t}\,e^{-\frac{\Gamma^+_+}{2} t}-e^{i(k_0 -  \omega_{+r} +i\frac{\Gamma^+_+}{2})\,t}-e^{-i(k_0 + \omega_{+r} -i\frac{\Gamma^+_+}{2})\,t}}{(k_0 -  \omega_{+r} +i\frac{\Gamma^+_+}{2})(k_0 +  \omega_{+r} -i\frac{\Gamma^+_+}{2})} \,,\label{indirec}  \eea  the negative frequency contribution is obtained by $\omega_{+r}\rightarrow -\omega_{+r}$. We refer to these as \emph{indirect} terms. These feature complex poles at $k_0= \pm (\omega_{+r}-i\frac{\Gamma^+_+}{2})$, yielding   terms proportional to
\be \mathbb{P}_+(\omega_{+r})_{ac}\frac{\rho_{cd}(\pm \omega_{+r})}{2\omega_{+r}}\,\mathbb{P}_+(\omega_{+r})_{bd}\,(1-e^{\pm 2i\omega_{+r}}\,e^{-\Gamma^+_+ t})(1+2n(\omega_{+r})) \ll 1 \,,\label{indisublead} \ee
where we have used that in the narrow width approximation $\omega_{+r} \gg \Gamma^+_+$. These indirect terms are of second order in the couplings and therefore subleading with respect to the direct terms.

Finally consider the contribution of a positive frequency of one quasinormal mode in one bracket and a negative frequency of the \emph{other} mode in the other bracket. We refer to these as \emph{interference} terms.  For example consider the positive frequency mode $\omega_{+r}$ in the first bracket in (\ref{noiscor7}), and the negative frequency mode $-\omega_{-r}$ in the second bracket, the time integrals yield a term proportional to
\be \Bigg[\frac{e^{i(k_0 - \omega_{+r} +i\frac{\Gamma^+_+}{2} )\,t}-1}{i(k_0 -  \omega_{+r} +i\frac{\Gamma^+_+}{2} )}\Bigg]\,\Bigg[\frac{e^{-i(k_0 -  \omega_{-r}-i\frac{\Gamma^-_-}{2} )\,t}-1}{(-i)(k_0 -  \omega_{-r} -i\frac{\Gamma^-_-}{2})}\Bigg] \,,\label{intefer}  \ee a similar analysis as for the previous terms yields
the following leading order contribution to the correlation function
\be \mathbb{P}_+(\overline{\omega} )_{ac}\, \frac{\rho_{cd}(\overline{\omega} )}{\big[\omega_{+r}-\omega_{-r}-\frac{i}{2}(\Gamma^+_+ + \Gamma^-_-)\big] }\,\mathbb{P}_-(-\overline{\omega} )_{bd}\,\Big[1-e^{ i(\omega_{+r}-\omega_{-r})\,t}\,e^{-\frac{1}{2}(\Gamma^+_+ + \Gamma^-_-)\,t}\Big]\, (1+2n(\overline{\omega}))\,,\label{interf}  \ee these interference terms exhibit the \emph{quantum beats}, an interference phenomenon associated    with the difference in the (quasi) normal mode frequencies, similar to that in the expression (\ref{qbits}).

These results are general and highlight the perturbative and non-perturbative contributions to the correlation functions in the long time limit. Before we discuss the non degenerate and nearly-degenerate cases, it is convenient to compare the results above to the case of the equal time correlation functions of a free field theory in thermal equilibrium, which is given by
\be \langle \Phi_a(\vk,t)  \Phi_b(-\vk,t) \rangle =  \frac{ \delta_{ab}}{2\omega_a(k)}\,(1+ 2 n(\omega_a)) \,,\label{fricorr}\ee where we assumed uncorrelated fields, and the brackets stand for statistical averages in a thermal ensemble of uncorrelated fields.

It is then clear that the long time limits (\ref{direint},\ref{indisublead},\ref{interf}) all feature  exponential relaxation to a \emph{thermalized stationary state} with the asymptotic long time limit featuring the thermal factors $1+2 n(\omega)$ in terms of the real parts of the frequencies of the quasinormal modes (the imaginary parts yield subleading contributions). This is one of the important results of this study.

 Furthermore, all  feature \emph{off-diagonal} terms   which we identify  as \emph{coherence} because of the similarity with two-level systems\cite{zubairy}, as discussed above.  We refer to this phenomenon as bath-induced coherence because even if the fields are initially uncorrelated, their interaction with the bath induces off-diagonal terms   which survive  in the long time limit. Interference terms between the two different quasinormal modes leads to the approach to the stationary thermal state with quantum beats.   Thermalization, the emergence of off diagonal coherence in the long time limit and quantum beats from interference between the two different quasinormal modes are some of the main results of this study. In the previous section (\ref{sec:loy}) we highlighted that in the (LOY) theory, quantum beats emerged in the time evolution of the total \emph{population} (\ref{ocuK}) and by unitarity in the amplitudes of intermediate states or decay products,   (see eqn. (\ref{combo})). In the effective field theory these interference terms are explicit in the approach to the stationary thermal state of the \emph{correlation functions of the fields}, both the diagonal and off-diagonal (coherence) components displaying the quantum beats.

\textbf{Non-degenerate case:}

In the non-degenerate case, with $\omega_{+r}-\omega_{-r} \simeq \omega_1-\omega_2 \gg \Gamma$ the direct terms (\ref{direint}) are the leading ones. As shown by eqns. (\ref{Ppnd},\ref{Pmnd}) with  (\ref{obet1},\ref{ogam1}), in the non-degenerate case the off-diagonal components of the projection operators are of $\mathcal{O}(g^2)$  . Therefore, in this case, the correlation functions exhibit thermalization albeit with  a perturbatively small  coherence. Furthermore, because the interference terms are perturbatively small since $\rho/(\omega_1-\omega_2) \propto \Sigma \propto g^2$, the quantum beats in the approach to thermalization feature  small amplitudes. This is in agreement with the results obtained in ref.\cite{mesonmix} for the case of (ultralight) axion-pion mixing.

\textbf{Nearly degenerate case:}
 In the nearly degenerate case with $\omega_{+r}-\omega_{-r} \lesssim  \Gamma$ both the direct (\ref{direint}) and the interference terms (\ref{interf})  are of $\mathcal{O}(1)$, and \emph{all} the matrix elements of the projector operators are also of $\mathcal{O}(1)$. In this case the amplitude of the quantum beats is large, enhanced by the near resonance, and the time scale of these interference effects is  similar to the relaxation time scale. In this case the off diagonal correlations, namely the coherence becomes large, amplified by the
 (near) resonant denominators, and could potentially be observable. This situation is akin to the case of $K^{0}-\overline{K}^{(0)}$ mixing where the decay products exhibit quantum beats on the time scales comparable to the lifetime. This is clearly the same physical process as described by the (LOY) theory described in section (\ref{sec:loy}). However, in the effective field theory approach the quantum beats are explicit in the correlation functions of the mixing fields both in the diagonal and off-diagonal components, and in the approach to the thermal stationary state.

 This large amplitude interference effect may open a window towards observation of synthetic-cosmological axion mixing via their (anomalous) coupling to photons with a Chern-Simons term. This pathway is being explored as a possible mechanism to harness synthetic axion quasiparticles in condensed matter systems to probe the cosmological axion\cite{shuyboy}.

\subsection{Relation to the (LOY) formulation of mixing:}\label{subsec:wwcompa}

The results of the effective field theory  bear a similarity with those obtained from the (LOY) theory in section (\ref{sec:loy}), but also noteworthy differences.   We seek to  establish  a more direct correspondence between both formulations     enlightening  the reason for the similarities and the origin of the differences.

The main ingredient to obtain the time evolution of expectation values and correlation functions is the Green's function (\ref{gomegt}) which is completely determined by the solutions of the Langevin equation (\ref{langevin}) for the homogeneous case $\xi_a\equiv 0$, namely
 \begin{equation}
        \ddot{\Phi}_a(\vec{k};t) + \omega^2_a( {k})\, \Phi_a(\vec{k};t) + \int_0^t \Sigma_{ab}(\vec{k};t-t') \Phi_b(\vec{k};t') dt' = 0 \,.\label{homolang}
    \end{equation}
In absence of the self-energy, the solutions are the usual free field positive and negative frequency components with constant amplitudes. Since the self-energy is $\propto g^2$ (with $g$ a generic coupling), we write
\be \Phi_a(\vec{k};t) \equiv \C_a(\vec{k};t)\,e^{-i \omega_a( {k})\,t}+\C^*_a(\vec{k};t)\,e^{i \omega_a( {k})\,t} \,,\label{amps}\ee  where the amplitudes $\C_a(\vec{k};t), \C^*_a(\vec{k};t)$ are slowly varying, namely $\dot{\C_a}, \dot{\C}^*_a \propto \Sigma \propto g^2$.  The equations of motion (\ref{homolang}) become
\bea e^{-i \omega_a( {k})\,t}\Bigg[\ddot{\C_a}(\vec{k},t)-2i\omega_a( {k})\,\dot{\C_a}(\vec{k},t)   + e^{-i(\omega_b( {k})-\omega_a( {k}))t }\int_0^t \Sigma_{ab}(\vec{k};t-t')\,e^{i\omega_b( {k})\,(t-t')}\, \C_b(\vec{k};t') dt' \Bigg] + \nonumber \\
e^{i \omega_a( {k})\,t}\Bigg[\ddot{\C}^*_a(\vec{k},t)+2i\omega_a( {k})\,\dot{\C}^*_a(\vec{k},t)   + e^{i(\omega_b( {k})-\omega_a( {k}))t }\int_0^t \Sigma_{ab}(\vec{k};t-t')\,e^{-i\omega_b( {k})\,(t-t')}\, \C^*_b(\vec{k};t') dt' \Bigg] =0 \nonumber \\ \,,\label{slowvars}
\eea where in the last terms the sum over $b$ is implicit. Because the terms inside the brackets are slowly varying and of $\mathcal{O}(g^2)$, each bracket must vanish independently, yielding
\be \ddot{\C_a}(\vec{k},t)-2i\omega_a( {k})\,\dot{\C_a}(\vec{k},t)   + e^{-i(\omega_b( {k})-\omega_a( {k})t) }\int_0^t \Sigma_{ab}(\vec{k};t-t')\,e^{i\omega_b( {k})\,(t-t')}\, \C_b(\vec{k};t') dt' = 0 \,,\label{sloweqn}\ee the equation for $\C^*$ is obtained from (\ref{sloweqn}) by replacing $\omega_{a,b} \rightarrow -\omega_{a,b}$. Let us neglect $\ddot{\C}$ in (\ref{sloweqn}) for a moment, we will show below that it is subleading in the long time limit. Introduce
\be \mathcal{W}_{ab}[t;t'] = \frac{i}{2\omega_a( {k})}\,\int_0^{t'} \Sigma_{ab}(\vec{k};t-t'')\,e^{i\omega_b( {k})\,(t-t'')}\,   dt'' ~~;~~ \mathcal{W}_{ab}[t;0]=0\,,\label{bigW}\ee
in terms of which  (\ref{sloweqn}) becomes
\be \dot{\C_a}(\vec{k},t)  = -e^{-i(\omega_b( {k})-\omega_a( {k}))t }\int_0^t\,\Big(\frac{d}{dt'} \mathcal{W}_{ab}[t;t']\Big) \, \C_b(\vec{k};t') dt' \,.\label{nuslow}\ee Upon integration by parts and using the initial condition in (\ref{bigW}), the integral becomes
\be \int_0^t\,\Big(\frac{d}{dt'} \mathcal{W}_{ab}[t;t']\Big) \, \C_b(\vec{k};t') dt' =  \mathcal{W}_{ab}[t;t]\,\C_b(\vec{k};t) - \int_0^t\,  \mathcal{W}_{ab}[t;t'] ) \, \Big( \frac{d}{dt'}\C_b(\vec{k};t')\Big) dt' \,,\label{bypart}  \ee because $\mathcal{W} \propto g^2$ and $\dot{\C} \propto \Sigma \propto g^2$ the second term in eqn. (\ref{bypart}) is of $\mathcal{O}(g^4)$ and will be neglected to leading order, yielding
 \bea
\dot{\C}_1(\vec{k},t) & = & - \Big\{\mathcal{W}_{11}[t;t]\,\C_1(\vec{k};t)+  e^{i(\omega_1({k})-\omega_2({k}))t }\,\mathcal{W}_{12}[t;t]\,\C_2(\vec{k};t)\Big\}\,,\label{c1equ}\\
\dot{\C}_2(\vec{k},t) & = & -\Big\{e^{i(\omega_2({k})-\omega_1({k}))t }\,\mathcal{W}_{21}[t;t]\,\C_1(\vec{k};t)+   \mathcal{W}_{22}[t;t]\,\C_2(\vec{k};t)\Big\}\,,\label{c2equ}
\eea where
\be \mathcal{W}_{ab}[t;t]= \int^t_0 \int^{\infty}_{-\infty} \frac{dk_0}{(2\pi)}\,\frac{\rho_{ab}(k_0,k)}{2\omega_a(k)}\, e^{i(\omega_b(k)-k_0)(t-t')}\,dt' \,.\label{bigo}\ee

Comparing the amplitude equations \emph{for the positive frequency component} (\ref{c1equ},\ref{c2equ}) with the amplitude equations in the (LOY) formulation, eqns. (\ref{dotc1fin},\ref{dotc2fin}) with $W_{ab}[t,t]$ given by eqn. (\ref{Wabtt}), we see that they  are exactly the same with the identifications
\be \mathcal{W}_{ab}[t;t] \equiv W_{ab}[t;t]~~;~~ \mu_{ab}(k_0) \equiv \frac{\rho_{ab}(k_0,k)}{4\pi\omega_a(k)}~~;~~ E_{1,2} \equiv \omega_{1,2}(k) \,.\label{eqimu}\ee  Furthermore, this analysis clarifies that the amplitudes $\mathcal{C}^*_{a}$ for the negative frequency components   are also present in the effective field theory framework, but not in the (LOY) theory.

Invoking the long time limit
\be \int^{t}_0    e^{i(\omega_b(k)-k_0)(t-t' )}\,dt'~{}_{\overrightarrow{t\rightarrow \infty }}~ i\,\Bigg[\mathcal{P}\,\Big(\frac{1}{(\omega_b(k)-k_0)} \Big) -i \pi \delta(\omega_b(k)-k_0) \Bigg]\,,\label{lot} \ee yields
\be -i \mathcal{W}_{ab}[t;t]~ \rightarrow ~ \Delta_{ab}(\omega_b(k))\,,  \label{lotW}\ee where $\Delta_{ab}(\omega)$ is given by eqn. (\ref{sigofnucc}) with the identification (\ref{eqimu}). Since in the long time limit  $\mathcal{W}_{ab}[t;t]~ \rightarrow ~ \Delta_{ab}(\omega_b(k))$ it follows from the amplitude equations
(\ref{dotc1fin},\ref{dotc2fin}) that $\ddot{\mathcal{C}}_a \propto g^4$ and can be consistently neglected, thus justifying neglecting $\ddot{\mathcal{C}}$ in eqn. (\ref{sloweqn}). The equations for the amplitudes (\ref{c1equ},\ref{c2equ}) become exactly the same as the set of equations (\ref{dotc1fin},\ref{dotc2fin}) in the (LOY) theory, therefore, the positive frequency components of   the Green's function $\mathcal{G}_{ab}(k;t)$ eqn. (\ref{Gtotnm}), is equivalent to the bracket in eqn. (\ref{solLT}), explaining the similar projector operators. However, the full Green's function (\ref{Gtotnm})  includes the negative frequency components, because (\ref{Gtotnm})    describes the time evolution of fields rather than single particle amplitudes.

  The solutions of the Langevin equation that determine the expectation values and correlation functions in the effective field theory, namely (\ref{Asplits}) features \emph{two} terms. The homogeneous term  ($  \Phi^{h}_a(\vk;t)$ ) in (\ref{realtisol}) depends on the initial conditions and corresponds to    the solution (\ref{solLT}) in the (LOY) theory, which also depends on the initial conditions. However the inhomogeneous term  ($  \Phi^{\xi}_a(\vk;t)$ ) in (\ref{realtisol}) is independent of initial conditions and determined by the noise. It is this inhomogeneous term that determines the asymptotic behavior of the correlation functions and exhibits the approach to a thermal stationary state in the long time limit, while the homogeneous term decays exponentially at long time, in the same manner as the amplitudes in the (LOY) theory. This is one of the major differences between the effective field theory and (LOY) theory of mixing.

This analysis highlights the similarities and differences between the (LOY) theory and the effective field theory, the differences are noteworthy: \textbf{i:)}
the effective field theory describes the evolution of fields, including both positive and negative frequency components of the quasinormal modes, \textbf{ii:)}  the effective field theory description yields the correlation functions, describes the approach to a thermal steady state and  the emergence and long-time survival of coherence, aspects that are not captured by the (LOY) theory. Another important difference is that in the effective field theory,  the quantum beats are manifest in the approach to thermalization of the correlation function as a consequence interference of quasinormal modes, both in the diagonal (populations) and off-diagonal (coherence) components of the correlation functions. \textbf{iii:)} Since the (LOY) method describes the evolution of the amplitudes of pure, single particle states, it cannot describe correlation functions.

\vspace{1mm}

\section{Summary of results and conclusions:}\label{sec:conclusions}

\textbf{Summary of results:}
   We  generalized the seminal theory of particle mixing pioneered by Lee, Oehme and Yang (LOY) to study CP violation in $K^0-\overline{K}^0$ mixing. This theory is the cornerstone of all analysis of CP violation in flavored meson mixing in terms of an effective non-hermitian Hamiltonian.

   We extend this theory in two ways: i) to include    the cases in which the mixing degrees of freedom are not mass-degenerate in absence of perturbations, thereby relaxing the assumption of CPT invariance, ii) we treat the time evolution without resorting to the approximation of a time independent non-Hermitian effective Hamiltonian, and discuss the caveats resulting from this approximation, which become more important in the non-degenerate case.  The (LOY) theory   is only valid for pure single (or few) particle states and does not directly allow to obtain correlation functions of the mixing fields, nor the time evolution of multiparticle states, such as coherent states, or statistical ensembles. However, its generalization and extension provides a useful guide to and benchmark for the effective field theory which  we introduce to describe \emph{indirect} particle mixing   as a consequence of their coupling to a common set of intermediate states or decay channels populated in a \emph{medium}.

   The effective action determines the time evolution of the \emph{reduced} density matrix after tracing over the degrees of freedom in the medium described as a bath in thermal equilibrium. Therefore, it describes the dynamics of field mixing as a quantum open system. Indirect mixing is a result of non-vanishing correlations of the operators that couple the mixing partners to the intermediate
states in the medium, and is manifest in off-diagonal components of the self-energy. The dynamics of field mixing is determined by a Langevin-like equation of motion with a dissipative self-energy kernel and stochastic noise obeying a generalized fluctuation dissipation relation.  The  solution of the equations of motion determine the dynamics of expectation values and correlation functions in terms of a superposition of quasinormal modes in the medium. The off-diagonal elements of the self-energy and noise kernels lead to indirect mixing and   the emergence of long-lived coherence, namely off-diagonal components of the two point correlation functions, even when initially the mixing fields are uncorrelated. We  refer to this phenomenon as \emph{bath induced coherence}.  We analyze in detail the cases in which the masses of the mixing particles are widely different, namely the non-degenerate case, and when they are nearly degenerate, which \emph{may} describe small violations of CPT. In both cases even if one of the fields features an initial expectation value and the other does not, the latter develops an expectation value as a consequence of indirect mixing. We find the remarkable result that the equal time two point correlation functions of the fields approach a \emph{thermal stationary state} and feature \emph{quantum beats} as a consequence of the interference of the quasinormal modes.   In the non-degenerate case these interference effects feature perturbatively small amplitudes, however, in the non-degenerate case the amplitude of the quantum beats is   resonantly enhanced and non-perturbative. These interference effects \emph{may} provide an observational avenue to probe cosmological axions in condensed matter systems.

We establish a direct relation between the effective field theory and the (LOY) theory of mixing, and highlight important differences, in particular that the effective field theory describes emergent, bath induced  long-lived coherence independent of the initial conditions that approach asymptotically a stationary thermal state.

\vspace{1mm}

\textbf{Conclusions:}

Indirect field mixing as a consequence of common intermediate states or decay channels is of great importance in particle physics, cosmology and possibly condensed matter physics. In particle physics indirect field mixing is at the heart of flavor meson mixing and CP violation in the standard model. Beyond the standard model  it may be a consequence of intermediate messengers connecting standard model particles to degrees of freedom beyond through ``portals''. In cosmology  various axion-like particles may mix through common decay channels into photons and or gluons, and in condensed matter    ``synthetic'' axions, emergent quasiparticles in materials that feature parity breaking, such as topological insulators and Weyl semimetals  may hybridize (mix) with cosmological axions thereby offering a way to probe the latter by exciting the former. Thus, the interdisciplinary relevance  of field mixing motivates the study in this article.
An important result of this study is that the equal time correlation functions feature quantum beats, as a consequence of interference of the quasinormal modes in the medium.

As demonstrated within  the (LOY) theory, quantum beats are also manifest in the time evolution of the decay products, which may provide an observational signature of field mixing. This could be of particular relevance in the case of axion mixing.

The phenomena revealed by this study, such as bath induced emergent coherence, induced condensates and quantum beats, are all qualitatively general independent of the particular couplings or
degrees of freedom in the medium. However, the quantitative form of the quasinormal modes, the projection operators and amplitudes of the quantum beats clearly will depend on the particular models and the parameters that define them.

Although we focused in field mixing in the case of bosonic fields, the general approach is also suitable  to study indirect mixing for fermionic or gauge degrees of
freedom. In the case of fermions the derivation of the effective field theory would require the extension of the current study to Grassman fields. One possible avenue would be to study neutrino mixing in the \emph{mass basis}, where the weak interaction vertices feature flavor-off diagonal terms after diagonalizing a mass matrix in the free part of the Lagrangian. An effective field theory description of indirect mixing (of the mass eigenstates) in a medium in which vector bosons and charged leptons are in thermal equilibrium may be a suitable application of the concepts developed in this study that may be worthwhile to study further. The effective field theory approach may complement the study of neutrinos\cite{kainu} and axions  in a medium including condensates\cite{aimarsh} with kinetic or Boltzman equations and allow to obtain off-diagonal correlation functions, namely coherences,  not just populations.  We expect many features of the results found in this study to be common to other field-mixing scenarios, for example we conjecture that  the emergence of long-lived coherence (off diagonal correlation functions), approach to thermalization  and quantum beats, as a result of the interference between (quasi)-normal modes in the medium are robust consequences of field mixing that may yield to novel phenomena, and plausible observational consequences, worthy of further exploration.

Among further questions that remain to be addressed in future studies are the issues of renormalizability, in particular if the off diagonal matrix elements of the self-energy
feature divergences, renormalizing them may necessitate off diagonal counterterms in the bare Lagrangian. This would call for \emph{direct} mixing terms (such as an off diagonal mass matrix) to be included in the bare Lagrangian. These aspects must be studied on a model dependent basis, since the renormalization aspects are directly related to the
type of operators $\mathcal{O}_a[\chi]$. Furthermore, we have \emph{assumed} that $\langle \mathcal{O}[\chi] \rangle =0$, however a non-vanishing expectation value of this operator in the medium would require introducing tadpole terms that may lead to condensates of the fields $\phi_a$. All of these questions while interesting in their own right, remain for further study.

\acknowledgements
  The authors thank Wen-Yuan Ai, Ankit Beniwal, Angelo Maggi and David Marsh for stimulating and enlightening discussions,  and gratefully acknowledge  support from the U.S. National Science Foundation through grant   NSF 2111743.

\appendix

\section{Single species}\label{app:single}
In this appendix we gather the results of the Weisskopf-Wigner approximation in the simpler case of one species to highlight the main aspects associated with the fulfillment of unitarity and the differences between the exact results via Laplace transform and the Markov approximation.
For a single species, $\phi$  we have \bea \dot{C}_\phi(t) & = & -i \sum_{\kappa} \langle \phi|H_I(t)|\kappa\rangle \,C_\kappa(t)\label{CA}\\
\dot{C}_{\kappa}(t) & = & -i \, C_\phi(t) \langle \kappa|H_I(t) |\phi\rangle \label{Ckapas}\eea where the sum over $\kappa$ is over all the intermediate states coupled to $|\phi\rangle$ via $H_I$.

Consider the initial value problem in which at time $t=0$ the state of the system $|\Psi(t=0)\rangle = |\phi\rangle$ i.e. \be C_\phi(0)= 1,\   C_{\kappa}(0) =0 .\label{appinitial}\ee  We can solve eq.(\ref{Ckapas}) and then use the solution in eq.(\ref{CA}) to find \bea  C_{\kappa}(t) & = &  -i \,\int_0^t \langle \kappa |H_I(t')|\phi\rangle \,C_\phi(t')\,dt' \label{Ckapasol}\\ \dot{C}_\phi(t) & = & - \int^t_0 \sigma(t-t') \, C_\phi(t')\,dt' \label{intdiffapp} \eea where
\be \sigma(t-t') = \sum_\kappa |\langle \phi|H_I |\kappa\rangle|^2 e^{i(E_\phi-E_\kappa)(t-t')} \equiv \int^\infty_{-\infty} \mu(k_0)\,e^{-i(k_0-E_\phi)(t-t')}\,dk_0 \label{sigma} \ee
and we introduced the spectral density
\be \mu(k_0) = \sum_{\kappa}|\langle \phi|H_I |\kappa\rangle|^2\,\delta(k_0-E_\kappa)\,.,\label{muspec}\ee
  Inserting the solution for $C_\phi(t)$ into eq.(\ref{Ckapasol}) one obtains the time evolution of amplitudes $C_{\kappa}(t)$ from which we can compute $|C_\kappa(t)|^2$, namely,  the time dependent probability to populate the state $|\kappa\rangle$,  .

The set of equations (\ref{CA},\ref{Ckapas}), together with the hermiticity of the interaction Hamiltonian $H_I$,   yields
\be \frac{d}{dt} \Big[|C_\phi(t)|^2+\sum_{\kappa} |C_\kappa(t)|^2   \Big]=0 \,  \label{timder1}\ee which along with the initial conditions (\ref{appinitial}) lead to the unitarity relation
\be \Big[|C_\phi(t)|^2+\sum_{\kappa} |C_\kappa(t)|^2   \Big] =1\,.\label{appunita}\ee

\vspace{1mm}

\textbf{Exact solution of eqn. (\ref{intdiffapp})}
The integro-differential equation (\ref{intdiffapp}) for $C_\phi(t)$ can be solved by Laplace transform. Introducing the Laplace variable $s$ and the Laplace transform of $C_\phi(t)$ as $\mathcal{C}_\phi(s)$, with the initial condition $C_\phi(t=0)=1$, we find \be \mathcal{C}_\phi(s)= \Bigg[s+\int_{-\infty}^\infty dk_0 ~ \frac{\mu(k_0)}{s+i(k_0'-E_\phi)}\Bigg]^{-1} \label{Lapla} \ee with solution \be C_\phi(t) = \int^{i\infty+\epsilon}_{-i\infty +\epsilon} \frac{ds}{2\pi\,i} ~\mathcal{C}_\phi(s)\,e^{st} \label{invlapla}\ee where the $\epsilon \rightarrow 0^+$ determines the Bromwich contour in the complex $s$-plane parallel to the imaginary axis to the right of all the singularities. Writing $s=i(-\omega-i\epsilon)$ we find \be C_\phi(t) = -\int_{-\infty}^{\infty} \frac{d\omega}{2\pi\,i}~ \frac{e^{-i\omega t}}{\Big[\omega- \int_{-\infty}^{\infty} d\omega'~\frac{\mu(k_0)}{E_\phi+\omega-k_0+i\epsilon}+i\epsilon \Big]  }\,.\label{CAfin}\ee The integral is carried out by closing the contour in the lower half $\omega$-plane. In the free case where $\mu(k_0) =0$, the pole is located at $\omega =-i\epsilon \rightarrow 0$, leading to a constant $C_\phi$. In perturbation theory there is a complex pole very near $\omega =0$ which can be obtained directly by
expanding the integral in the denominator near $\omega =0$. We find \be  \int_{-\infty}^{\infty} dk_0'~\frac{\mu(k_0)}{E_\phi+\omega-k_0+i\epsilon} \simeq  \Delta E_\phi - z_\phi\,\omega - i \,\frac{\Gamma_\phi}{2} \label{aproxi}\ee where \bea \Delta E_\phi  & = & \mathcal{P} \int_{-\infty}^{\infty} d\omega' \, \frac{\mu(k_0)}{(E_\phi-k_0)} \label{energyshift} \\ \Gamma_\phi & = & 2\pi\,\mu(E_\phi) \label{width} \\ z_\phi & = & \mathcal{P} \int_{-\infty}^{\infty} dk_0 \, \frac{\mu(k_0)}{(E_\phi-k_0)^2}\label{smallz}\eea and $\mathcal{P}$ stands for the principal part. The term $\Delta E_\phi$ is recognized as the energy shift while $\Gamma_\phi$ is seen to be the decay rate as found from Fermi's golden rule. The {\em long time} limit of $C_\phi(t)$ is determined by this complex pole near the origin leading to the asymptotic behavior to leading order in the coupling \be C_\phi(t)\simeq  {Z}_\phi \, e^{-i\Delta E_\phi\,t}\,e^{-\frac{\Gamma_\phi}{2}\,t} \label{tasi}\ee where
\be  {Z}_\phi = \frac{1}{1+z_\phi}\simeq 1-z_\phi = \frac{\partial}{\partial E_\phi} \big[ E_\phi + \Delta E_\phi \big] \label{wavefunc}\ee is the wave function renormalization constant.

\vspace{1mm}

\textbf{Markov approximation.}

The time evolution of $C_\phi(t)$ determined by eq.(\ref{intdiffapp}) is \emph{slow} in the sense that
the time scale is determined by a weak coupling kernel $\sigma\propto H^2_I$. This suggests to use a Markovian approximation in terms of a
consistent expansion in derivatives of $C_\phi$. For this purpose, let us define \be W (t,t') = \int^{t'}_0 \sigma(t-t'')dt'' \label{Wo}\ee so that \be \sigma(t-t') = \frac{d}{dt'}W (t,t'),\quad W (t,0)=0. \label{rela} \ee Integrating by parts in eq.(\ref{intdiffapp}) we obtain \be \int_0^t \sigma(t-t')\,C_\phi(t')\, dt' = W(t,t)\,C_\phi(t) - \int_0^t W (t,t')\, \frac{d}{dt'}C_\phi(t') \,dt'. \label{marko1}\ee The second term on the right hand side is formally of \emph{fourth order} in $H_I$ because $W(t,t') \simeq H^2_I$ and $\dot{C}_\phi(t) \simeq H^2_I$, therefore it can be neglected to leading order  $\mathcal{O}(H^2_I)$.  Up to leading order in this Markovian approximation the equation eq.(\ref{intdiffapp}) becomes \be \dot{C}_\phi(t) + W_0(t,t) C_\phi(t) =0\,, \label{markovian}\ee with the solution \be C_\phi(t) = e^{-i\int_0^t \mathcal{E}(t')dt'},\quad \mathcal{E}(t) =
-i\,W (t,t) \,.\label{solumarkov}\ee Note that in general $\mathcal{E}(t)$ is complex.   To leading order in $H^2_I$ we find
\be \mathcal{E}(t) = -i\int^t_0 \sigma(t-t')\,dt' =  \int_{-\infty}^{\infty} dk_0 \, \frac{\mu(k_0)}{( E_\phi-k_0)}\,\Big[ 1-e^{-i(k_0-E_\phi)t} \Big]\label{Wominko} \ee  so that \bea \int^t_0 \mathcal{E}(t')\,dt' & = &
 t\,\int_{-\infty}^{\infty} dk_0 \, \frac{\mu(k_0)}{( E_\phi-k_0)}\,\Big[ 1-\frac{\sin(k_0-E_\phi)t}{(k_0-E_\phi)t} \Big] \nonumber \\ & - & i  \int_{-\infty}^{\infty} dk_0 \, \frac{\mu(k_0)}{( E_\phi-k_0)^2}\,\Big[ 1-\cos\big[(k_0-E_\phi)t\big] \Big] \label{energyminko}\eea

Asymptotically as $t\rightarrow \infty$, these integrals approach:
 \be \int_{-\infty}^{\infty} dk_0 \, \frac{\mu(k_0)}{( E_\phi-k_0)}\,\Big[ 1-\frac{\sin(k_0-E_\phi)\,t}{(k_0-E_\phi)\,t} \Big] ~~{}_{\overrightarrow{t\rightarrow \infty}} ~~ \mathcal{P}\int_{-\infty}^{\infty} dk_0 \, \frac{\mu(k_0)}{( E_\phi-k_0)} \label{realpartofE}\ee
 \be \int_{-\infty}^{\infty} dk_0 \, \frac{\mu(k_0)}{( E_\phi-k_0)^2}\,\Big[ 1-\cos\big[(k_0-E_\phi)t\big] \Big]~~{}_{\overrightarrow{t\rightarrow \infty}} ~~\pi\,t\, \mu(E_\phi) + \mathcal{P} \int_{-\infty}^{\infty} dk_0 \, \frac{\mu(k_0)}{( E_\phi-k_0)^2} \label{imagE}\ee Using these results we find that in the long time limit,
  \be -i \int^{t}_0 \mathcal{E}(t')\,dt' \rightarrow -i \Delta E_\phi~t - \frac{\Gamma_\phi}{2}~t -z_\phi  \,, \label{phaseminko} \ee where $\Delta E_\phi,\Gamma_\phi,z_\phi$ are given by eqs. (\ref{energyshift}, \ref{width},\ref{smallz},\ref{wavefunc}). From this we read off \be C_\phi(t) = {Z}_\phi\,e^{-i\Delta E_\phi~t}\,e^{-\frac{\Gamma_\phi}{2}~t} \label{CAmarkovres}\ee  where we approximated $e^{-z_\phi } \simeq 1-z_\phi=Z_\phi$ up to second order in perturbation theory. This is in complete agreement with the asymptotic result from the exact solution eq.(\ref{tasi}) obtained via the Laplace transform.

  \vspace{1mm}

  \textbf{Taking the long time limit before integration:} We now compare the results obtained above with those obtained with yet another approximation: taking the long time limit in $W(t,t)$ in (\ref{markovian}) \emph{before} integrating this  evolution equation.
  \be W(t,t)_{\overrightarrow{t \rightarrow \infty}}  \int^{\infty}_{-\infty} \mu(k_0) \int^{\infty}_0 e^{-i(k_0-E_\phi)\tau} \,d\tau\,dk_0 =  i \Big\{   \int^{\infty}_{-\infty} \mathcal{P} \,  \frac{\mu(k_0)}{(E_\phi-k_0)}\,dk_0 - i\pi \mu(E_\phi)   \Big\}\,,\label{infiti} \ee therefore under this approximation the solution of (\ref{markovian}) is
  \be C_\phi(t) = e^{-i\Delta E_\phi~t}\,e^{-\frac{\Gamma_\phi}{2}~t} \,.\label{infilim}\ee Obviously, the main difference with the solutions (\ref{tasi},\ref{markovian}) is the lack of wave function renormalization in (\ref{infilim}). Therefore we conclude that the Markov approximation leading to (\ref{markovian}) reproduces the exact result obtained from Laplace transform, however the further approximation of replacing $W(t,t)$ by its infinite time limit (\ref{infiti}) in the Markovian equation (\ref{markovian}) misses the wavefunction renormalization.

\vspace{1mm}

\textbf{Unitarity:} Because of the exponential decay of the amplitude of the initial state, the unitarity condition (\ref{appunita}) entails that in the long time limit
\be \sum_{\kappa}|C_{\kappa}(\infty)|^2 =1\,. \label{infck}\ee We now address how this constraint is fulfilled. The coefficients $C_\kappa(t)$ are given by eqn. (\ref{Ckapasol}).

Introducing the leading order result (\ref{infilim}) (since $Z_\phi = 1 + \mathcal{O}(H^2_I)$)  into eq.(\ref{Ckapasol}) for the coefficients $C_\kappa$ we find to leading order
 \be |C_\kappa(\infty)|^2 = \frac{|\langle \kappa|H_I|\phi\rangle|^2}{\Bigg[(E^R_\phi -E_\kappa)^2 + \frac{\Gamma^2_\phi}{4} \Bigg]},\quad E^R_\phi = E_\phi+\Delta E_\phi. \label{populationkapa}\ee This expression can be interpreted as follows. If $|\phi\rangle$ is an unstable state,  the states $|\kappa\rangle$ with $E_\kappa \sim E^R_\phi$ i.e. those nearly resonant with the state $|\phi\rangle$, are ``populated'' with an amplitude $\propto 1/\Gamma_\phi$ within a band of width $\Gamma_\phi$ centered at $E^R_\phi$. Furthermore
 \be \sum_{\kappa} |C_{\kappa}(\infty)|^2  = \int_{-\infty}^{\infty}  \, \frac{\mu(k_0)}{\Big[(E^R_\phi -k_0)^2 + \frac{\Gamma^2_\phi}{4} \Big]}\,dk_0 \simeq 1 \label{unitaritydecay} \ee where we  have written
 \be \frac{1}{\Bigg[(E^R_\phi -k_0)^2 + \frac{\Gamma^2_\phi}{4} \Bigg]} = \frac{1}{\Gamma_\phi}\,\frac{\Gamma_\phi}{\Big[(E^R_\phi -k_0)^2 + \frac{\Gamma^2_\phi}{4} \Big]}\,,\ee    in the narrow width limit $\Gamma_\phi \rightarrow 0$ we replace
 \be \frac{\Gamma_\phi}{\Big[(E^R_\phi -k_0)^2 + \frac{\Gamma^2_\phi}{4} \Big]} \rightarrow 2\pi\,\delta(E^R_\phi -k_0)\,, \ee and used the result (\ref{width}) to obtain (\ref{unitaritydecay}).
  Unitarity entails a \emph{probability flow} from the initial towards the final excited states.

\section{Lehmann representation of correlation functions}\label{app:spectralrep}
The correlation functions $G^>_{ab}(x-y);G^<_{ab}(x-y)$ can be written in an exact Lehmann (spectral) representation which is useful to include in the equations of motion.
\bea G^>_{ab}(x-y) & = & \frac{1}{Z_\chi}\mathrm{Tr} e^{-\beta H_{\chi}} \mathcal{O}_a(x)\mathcal{O}_b(y)\, \label{ggreatap} \\ G^<_{ab}(x-y) & = & \frac{1}{Z_\chi}\mathrm{Tr} e^{-\beta H_{\chi}} \mathcal{O}_b(y)\mathcal{O}_a(x)\,, \label{glessap}\eea
where $Z_{\chi} = \mathrm{Tr} e^{-\beta H_{\chi}}$ and
$\mathcal{O}_{a}(\vx,t)= e^{iH_{\chi}t}\,e^{-i\vec{P}\cdot \vx}\, \mathcal{O}_{a}(0)\,e^{i\vec{P}\cdot \vx}\,e^{-iH_{\chi}t}$. In terms of a complete set of simultaneous eigenstates of $H_{\chi},\vec{P}$, namely $(H_{\chi},\vec{P})|n\rangle = (E_n,\vec{P}_n)|n\rangle $ and inserting the identity in this basis, we find
\bea   G^>_{ab}(x_1-x_2)  &  = &   \frac{1}{Z_{\chi}}{\sum_{n,m}} e^{-\beta E_n} e^{i(E_n-E_m)(t_1-t_2)}\,e^{-i(\vec{P}_n-\vec{P}_m)\cdot(\vx_1-\vx_2)}\,\langle n|\mathcal{O}_a(0)|m\rangle \langle m|\mathcal{O}_b(0)|n\rangle \,, \nonumber \\\label{ggreatrep}\\
G^<_{ab}(x_1-x_2)  &  = &   \frac{1}{Z_{\chi}}{\sum_{n,m}} e^{-\beta E_n} e^{-i(E_n-E_m)(t_1-t_2)}\,e^{i(\vec{P}_n-\vec{P}_m)\cdot(\vx_1-\vx_2)}\,\langle n|\mathcal{O}_b(0)|m\rangle \langle m|\mathcal{O}_a(0)|n\rangle \,. \nonumber \\ \label{glessrep}
\eea
 These representations may be written in terms of spectral densities, by introducing
 \bea && \rho^>_{ab}(k_0,\vk)  =   \frac{(2\pi)^4}{Z_{\chi}}{\sum_{n,m}} e^{-\beta E_n}  \,\langle n|\mathcal{O}_a(0)|m\rangle \langle m|\mathcal{O}_b(0)|n\rangle \, \delta(k_0 - (E_m-E_n)) \delta^{3}(\vk- (\vec{P}_m-\vec{P}_n))  \,, \nonumber \\ \label{rhogreatrep}\\
&& \rho^<_{ab}(k_0,\vk)     =     \frac{(2\pi)^4}{Z_{\chi}}{\sum_{n,m}} e^{-\beta E_n}  \,\langle n|\mathcal{O}_b(0)|m\rangle \langle m|\mathcal{O}_a(0)|n\rangle  \delta(k_0 - (E_n-E_m)) \delta^{3}(\vk- (\vec{P}_n-\vec{P}_m))\,,\nonumber  \\ \label{rholessrep}
\eea in terms of which
\bea G^>_{ab}(x_1-x_2)  &  = & \int \frac{d^4k}{(2\pi)^4}\,  \rho^>_{ab}(k_0,\vk)\, e^{-ik_0(t_1-t_2)}\,e^{i\vk\cdot(\vx_1-\vx_2)} \,\label{specgreat} \\
 G^<_{ab}(x_1-x_2)  &  = & \int \frac{d^4k}{(2\pi)^4}\,  \rho^<_{ab}(k_0,\vk)\, e^{-ik_0(t_1-t_2)}\,e^{i\vk\cdot(\vx_1-\vx_2)} \,.\label{specless} \eea Relabelling $n \leftrightarrow m$ and using the $k_0$ delta function   in (\ref{rholessrep}), we find the generalized Kubo-Martin-Schwinger condition\cite{kms}
 \be  \rho^<_{ab}(k_0,\vk) = e^{-\beta k_0}\, \rho^>_{ab}(k_0,\vk)\,. \label{kmscond} \ee Introducing the spectral density
 \be  \rho_{ab}(k_0,\vk)=  \rho^>_{ab}(k_0,\vk)- \rho^<_{ab}(k_0,\vk)= \rho^>_{ab}(k_0,\vk)\,\big(1-e^{-\beta k_0}\big) \,,\label{spectral}\ee it follows that
 \be \rho^>_{ab}(k_0,\vk)= \big(1+ n(k_0) \big) \rho_{ab}(k_0,\vk)~~;~~ \rho^<_{ab}(k_0,\vk)=  n(k_0) \, \rho_{ab}(k_0,\vk)\,,\label{relas}\ee
 where
 \be n(k_0) = \frac{1}{e^{\beta k_0}-1}\,. \label{nofk0}\ee

 Therefore the spatial Fourier transform of the self-energy matrix (\ref{kernelsigma}) and the noise kernel (\ref{kernelkappa})  can be written as
 \bea \Sigma_{ab}(k;t-t')& = & -i\int \frac{dk_0}{(2\pi)}\,\rho_{ab}(k_0,k)  e^{-ik_0(t-t')}  \label{appsigmadis} \\ \mathcal{N}_{ab}(k;t-t') & = & \frac{1}{2}\,\int \frac{dk_0}{(2\pi)}\,\rho_{ab}(k_0,k)\, \coth\big[ \frac{\beta k_0}{2}\big]\,  e^{-ik_0(t-t')} \,, \label{appnoisedis} \eea this is the general relation between the self-energy and the noise correlation function commonly determined by the spectral density $\rho_{ab}(k_0,k)$, a direct consequence of the fluctuation-dissipation relation as a result of the bath being in thermal equilibrium.

 Assuming rotational invariance implies that $\rho_{ab}(k_0,\vk) = \rho_{ab}(k_0,k)$, in particular the diagonal matrix elements of the spectral density
 \bea \rho_{aa}(k_0,k) & = &   \frac{(2\pi)^4}{Z_{\chi}}{\sum_{n,m}} e^{-\beta E_n}  \,|\langle n|\mathcal{O}_a(0)|m\rangle|^2 \, \big[ \delta(k_0 - (E_m-E_n)) - \delta(k_0 - (E_n-E_m))\big]\, \delta^{3}(\vk- (\vec{P}_m-\vec{P}_n)) \nonumber \\ &  = & -\rho_{aa}(-k_0,k)~~,~~\mathrm{no~ sum~over~a}\,. \label{oddro} \eea

The assumption of rotational invariance also applies to correlation functions of pseudoscalar operators (relevant for axions) in a thermal equilibrium density matrix which is invariant under rotations  , because these are bilinear in the operators, hence invariant under $\vk \rightarrow -\vk$.

We note that because the operators $\mathcal{O}_a$ are Hermitian, it follows that $(\rho^\gtrless_{ab}(k_0,k))^* = \rho^\gtrless_{ba}(k_0,k)$, consequently $\rho^*_{ab}(k_0,k) = \rho_{ba}(k_0,k)$.

 \vspace{1mm}

\section{Laplace Green's function}\label{app:green}
Consider the matrix
\be \mathbb{M} =  \left(
                                   \begin{array}{cc}
                                      M_{11} &  M_{12} \\
                                      M_{21} &  M_{22} \\
                                   \end{array}
                                 \right)\,,\label{mtxM}\ee

whose (right and left) inverse is
\be \mathbb{M}^{-1} = \frac{1}{det[\mathbb{M}]}\, \left(
                                   \begin{array}{cc}
                                      M_{22} &  -M_{12} \\
                                      -M_{21} &  M_{11} \\
                                   \end{array}
                                 \right)\,.\label{invmtxM}\ee In terms of the variables (\ref{mbar},\ref{D},\ref{abg}) it follows that
 \be \mathbb{M} =  \left(
                                   \begin{array}{cc}
                                      \overline{M}+ \frac{D}{2}\alpha &  \frac{D}{2}\beta \\
                                      \frac{D}{2}\gamma &  \overline{M}- \frac{D}{2}\alpha \\
                                   \end{array}
                                 \right)~~;~~ det[\mathbb{M}] = \Big(\overline{M}-\frac{D}{2}\Big)\Big(\overline{M}+\frac{D}{2}\Big)\,,\label{mtxM2}\ee    where we used the relation (\ref{abgrela}).

   Therefore the  inverse of the matrix (\ref{mtxM}) is given by
\be \mathbb{M}^{-1} = \frac{1}{\mathrm{det}[\mathbb{M}]  }\, \Bigg[ \overline{M} \, \mathbf{1}- \frac{ D}{2}\,\left(
                                                                           \begin{array}{cc}
                                                                             \alpha & \beta \\
                                                                             \gamma & -\alpha \\
                                                                           \end{array}
                                                                         \right) \Bigg]\,. \label{Gifi}\ee
 Writing
 \bea
 \overline{M}  & = &  \frac{1}{2}\,\Big(\overline{M}  + \frac{ D}{2} \Big)+ \frac{1}{2}\,\Big( \overline{M}  - \frac{ D}{2} \Big) \nonumber \\
  D & = & \Big( \overline{M}  + \frac{ D}{2} \Big)\,\Big( \overline{M}  - \frac{ D}{2} \Big) \, \Bigg[\frac{1}{\overline{M}  - \frac{ D}{2} }-\frac{1}{\overline{M}  + \frac{ D}{2} } \Bigg] \nonumber
 \eea   yields
 \be   \mathbb{M}^{-1} = \frac{\mathbb{P}_-}{\overline{M}  - \frac{ D}{2} } +  \frac{\mathbb{P}_+}{\overline{M}  + \frac{ D}{2} }~~;~~ \mathbb{P}_\pm = \frac{1}{2} \Big(\mathbf{1}\pm \mathbb{R} \Big) \,,\label{appgreenfin}   \ee with
 \be \mathbb{R} = \left(
                                                                           \begin{array}{cc}
                                                                             \alpha & \beta \\
                                                                             \gamma & -\alpha \\
                                                                           \end{array}
                                                                         \right)   ~~;~~ \mathbb{R}^2 = \mathbf{1}\,, \label{capR}\ee where the last equality follows from the identity (\ref{abgrela}), therefore the matrices $\mathbb{P}_\pm $ are projectors, namely
 \be   \mathbb{P}^2_\pm = \mathbb{P}_\pm \,,\label{projectors}\ee
           hence their eigenvalues are $0,1$.

\end{document}